\documentclass[11pt]{article}
\pdfoutput=1
\usepackage{amsmath,amssymb,epsfig,amsfonts}
\usepackage{graphicx,subfigure}
\usepackage[usenames, dvipsnames]{color}
\usepackage[pagebackref]{hyperref}
\usepackage{cite}
\usepackage{verbatim} 
\usepackage{float}
\restylefloat{table}
\usepackage[all]{xy}
\usepackage{pdflscape}
\usepackage{tikz}
\usepackage{array}
\usepackage{youngtab}
\usepackage{multirow}
\usepackage{color}
\usepackage{ulem}
\usepackage{tikz-cd}
\usepackage{bbold}
\usepackage{todonotes}

\usepackage{subfigure}

\DeclareFontFamily{U}{wncy}{}
    \DeclareFontShape{U}{wncy}{m}{n}{<->wncyr10}{}
    \DeclareSymbolFont{mcy}{U}{wncy}{m}{n}
    \DeclareMathSymbol{\Sh}{\mathord}{mcy}{"58} 

\makeatletter

\renewcommand*{\backref}[1]{}
\renewcommand*{\backrefalt}[4]{({%
    \ifcase #1 Not cited.%
          \or Page~#2.%
          \else Pages #2.%
    \fi%
    })}

\def\bbP{{\mathbb{P}}}
\def\bbZ{{\mathbb{Z}}}
\def\bbY{{\mathbb{Y}}}
\def\one{{\bf 1}}
\def\two{{\bf 2}}
\def\three{{\bf 3}}

\def\fku{{\mathfrak{u}}}

\def\fkg{{\mathfrak{g}}}

\newcount\hour \newcount\minute
\hour=\time \divide \hour by 60
\minute=\time
\def\now{%
\ifnum \hour<13
  \ifnum \hour=0 \advance \hour by 12 \number\hour:\else \number\hour:\fi%
     \ifnum \minute<10 0\fi%
     \number\minute%
\ A.M.%
\else \advance \hour by -12 \number\hour:%
  \ifnum \minute<10 0\fi%
  \number\minute%
  \ P.M.%
\fi%
}

\makeatother
\hypersetup{
    pdftitle={TASI Lectures on Abelian and Discrete Symmetries in F-theory},
    pdfauthor={Mirjam Cvetic, Ling Lin},
    pdfsubject={F-theory compactifications, TASI, abelian symmetries, Mordell--Weil, genus-one fibration, Tate--Shafarevich, Standard Model}
}

\begin{document}

\baselineskip=18pt  
\numberwithin{equation}{section}  
\allowdisplaybreaks  



\vspace*{-2cm} 
\begin{flushright}
{\tt UPR-1293-T}\\
\end{flushright}

\vspace*{0.8cm} 
\begin{center}
 {\LARGE  TASI Lectures on Abelian and Discrete Symmetries in F-theory}\\

 \vspace*{1.8cm}
 {Mirjam Cveti\v{c}$^{1,2,\dagger}$ and Ling Lin$^1$}\\

 \vspace*{1.2cm} 

{\it $^1$ Department of Physics and Astronomy, University of Pennsylvania,  \\
 Philadelphia, PA 19104-6396, USA}\\

 \bigskip
{\it $^2$ Center for Applied Mathematics and Theoretical Physics, \\
 University of Maribor, Maribor, Slovenia}\\

 \bigskip
  
 {{\tt cvetic@physics.upenn.edu}$\,,\quad$ {\tt lling@physics.upenn.edu}}

%
%
%
\vspace*{0.8cm}
\end{center}
\vspace*{.5cm}
%
\noindent
In F-theory compactifications, the abelian gauge sector is encoded in global structures of the internal geometry.
These structures lie at the intersection of algebraic and arithmetic description of elliptic fibrations:
While the Mordell--Weil lattice is related to the continuous abelian sector, the Tate--Shafarevich group is conjectured to encode discrete abelian symmetries in F-theory.
In these notes we review both subjects with a focus on recent findings such as the global gauge group and gauge enhancements.
We then highlight the application to F-theory model building.

\vfill

\begin{flushright}
	$^\dagger$ Lecturer at TASI 2017
\end{flushright}

\newpage

\tableofcontents
\newpage

\section{Introduction}

Over the past two decades, F-theory \cite{Vafa:1996xn, Morrison:1996na, Morrison:1996pp} has established itself as a powerful framework to study non-perturbative string compactifications.
A major part of its success is footed on the mathematical formulation of F-theory in terms of elliptic fibrations.
Utilizing tools from algebraic geometry, we have since learned about many intriguing connections between physics and mathematics.
A particularly active topic of research has been the understanding and systematic construction of abelian gauge symmetries in F-theory.
The original motivation arose from phenomenological considerations, where abelian symmetries were needed as selection rules in GUT model building \cite{Donagi:2009ra, Marsano:2009ym, Marsano:2009gv, Blumenhagen:2009yv,  Marsano:2009wr, Cvetic:2010rq, Grimm:2010ez, Weigand:2010wm, Dolan:2011iu, Marsano:2011nn, Krause:2011xj, Dolan:2011aq, Hayashi:2013lra,Cvetic:2013uta}.
In the absence of any direct detection of supersymmetry, it has further become more attractive to engineer the Standard Model gauge group directly, which of course relies on a realization of the hypercharge $U(1)$.
In addition, abelian symmetries provide novel links between physics and aspects of arithmetic geometry.

%

Unlike non-abelian symmetries, abelian ones are associated to inherently global data of the geometry.
In the case of continuous abelian symmetries, i.e., $U(1)$s, this geometric origin has been known since the early days of F-theory \cite{Morrison:1996pp}.
However, the first concrete global model with abelian symmetry, the so-called $U(1)$-restricted Tate model, was constructed much later \cite{Grimm:2010ez}.
This model explicitly realizes an elliptic fibration $\pi: Y \rightarrow B$ with a so-called rational section, which is essentially a copy of the base $B$ inside the total space $Y$ of the fibration.
Rational sections of elliptic fibrations form an abelian group---the famous Mordell--Weil group---which has been and still is the focus of many mathematicians.
It was not surprising that F-theory benefited immensely from their efforts.
Indeed, the introduction of the so-called Shioda-map to the F-theory community in \cite{Park:2011ji, Morrison:2012ei} sparked the explicit construction of many abelian F-theory models \cite{Mayrhofer:2012zy, Braun:2013yti, Borchmann:2013jwa, Cvetic:2013nia, Braun:2013nqa, Borchmann:2013hta,  Cvetic:2013uta, Cvetic:2013jta, Cvetic:2013qsa,Kuntzler:2014ila,  Klevers:2014bqa, Esole:2014dea, Lawrie:2014uya, Cvetic:2015ioa}.
The more formal approach to $U(1)$s via the Mordell--Weil group not only led to new insights about physical phenomena such as gauge symmetry breaking/enhancement or the global structure of the gauge group.
It also significantly improved the capabilities of F-theory model building (in addition to the previous references, see also \cite{Krippendorf:2014xba, Krippendorf:2015kta, Buchmuller:2017wpe}), which most recently culminated in globally consistent realizations of the chiral Standard Model spectrum \cite{Cvetic:2015txa, Lin:2016vus, Cvetic:2018ryq}.

The study of abelian symmetries also led to a drastic paradigm shift in the geometric description of F-theory.
Namely, it turned out that a consistent compactification space $Y$ need not to be elliptically fibered (i.e., having at least one rational section), but could more generally be a torus-, or genus-one fibration with a so-called multi-section \cite{Braun:2014oya}.
Physically, this reflects the presence of a gauged discrete abelian, i.e., $\bbZ_n$ symmetry, which can be viewed as the result of Higgsing a $U(1)$ with charge $n$ singlets \cite{Morrison:2014era, Anderson:2014yva, Garcia-Etxebarria:2014qua, Mayrhofer:2014haa, Klevers:2014bqa}.
Through duality to M-theory, $\bbZ_n$ symmetries are shown to be related to the so-called Tate--Shafarevich group $\Sh$ \cite{Braun:2014oya, Cvetic:2015moa}, which plays a role in arithmetic geometry of elliptic fibrations.
Though the full extend of the interplay between $\Sh$ and $\bbZ_n$ is not yet understood, the connection could possibly open up a physics-motivated method to construct examples of $\Sh$, which unlike the Mordell--Weil group is still quite mysterious in the mathematical literature.

Given the rich mathematical structures related to abelian symmetries in F-theory, these notes will provide a more formal approach to the topic.
After a brief introduction (section \ref{sec:basic_f-theory}) to F-theory, we will introduce in section \ref{sec:u1} the Mordell--Weil group, the Shioda-map, and their connection to $U(1)$ symmetries in F-theory.
There, we will also explain how these geometric objects encode the global gauge group structure of F-theory.
In section \ref{sec:discrete}, we then turn to discrete abelian symmetries and their dual descriptions in terms of multi-sections and torsional cohomology.
Finally, we reconnect these formal aspects to the original phenomenological motivations by presenting in section \ref{sec:application} three F-theory constructions that realize the gauge symmetry and the chiral spectrum of the Standard Model.
With the clear emphasis on abelian symmetries, many other detailed aspects of F-theory compactifications will be omitted or only highlighted briefly in section \ref{sec:outlook}.
For a more comprehensive review of F-theory, we refer to another set of TASI-lectures \cite{Weigand:2018rez}.
While these notes also include a detailed introduction to abelian symmetries, our presentation offers some complementary perspectives and puts the focus on some different aspects.

\section{Basics of F-theory Compactifications}\label{sec:basic_f-theory}

We will start with a brief recollection of F-theory compactification on elliptic fibrations, in order to make these notes self-contained.
For more details, we again refer to \cite{Weigand:2018rez}, and also to other reviews \cite{Denef:2008wq, Weigand:2010wm}.

To set the stage, we should first explain the central geometric object of F-theory, namely an an elliptically fibered Calabi--Yau manifold.
Such a space $Y_n \equiv Y$ is K\"ahler manifold of complex dimension $n$ with trivial first Chern class, together with a surjective holomorphic map $\pi: Y_n \rightarrow B_{n-1}$ onto a K\"ahler manifold $B_{n-1} \equiv B$ of complex dimension $n-1$.
The preimage $\pi^{-1}(p)$ of a generic point $p \in B_{n-1}$ is an elliptic curve with a marked point $O$, that is, a complex manifold of dimension 1 which is isomorphic to a torus $T^2$ with a distinguished origin.
As one varies the point $p$ along the base, the marked point $O$ varies holomorphically through $Y_n$, which defines the so-called zero section $\sigma_0: B_{n-1} \rightarrow Y_n$ of the elliptic fibration.
Being a holomorphic map from the base $B$ into the total space $Y$, its image defines a copy of the base, sitting as a divisor (a complex codimension one variety) of $Y$.

Any elliptic fibration can be described by a so-called Weierstrass model.
This description embeds the fiber as a curve inside a weighted projective surface $\bbP_{231}$ with projective coordinates $[x:y:z] \cong [\lambda^2 x : \lambda^3 y : \lambda z]$, cut out by the Weierstrass equation
\begin{align}\label{eq:weierstrass_equation}
	y^2 = x^3 + f\,x\,z^4 + g\,z^6 \, .
\end{align}
By promoting $f,g$ to functions over a base $B$, \eqref{eq:weierstrass_equation} then describes how the fiber varies over $B$, i.e., models the fibration $Y$.
The zero section $\sigma_0$ is described by the intersection with $z=0$, marking the point $O = [1:1:0]$ on each fiber.
One consistency condition of F-theory is that the elliptic fibration $Y$ is a Calabi--Yau space.
This is guaranteed, if the functions $f$ and $g$ are holomorphic sections of the line bundles ${\cal O}(K_B^{-4})$ and ${\cal O}(K_B^{-6})$, respectively, where $K_B$ is the canonical class of the base $B$.

\begin{figure}[!ht]
\centering
	\includegraphics[width=.9\hsize]{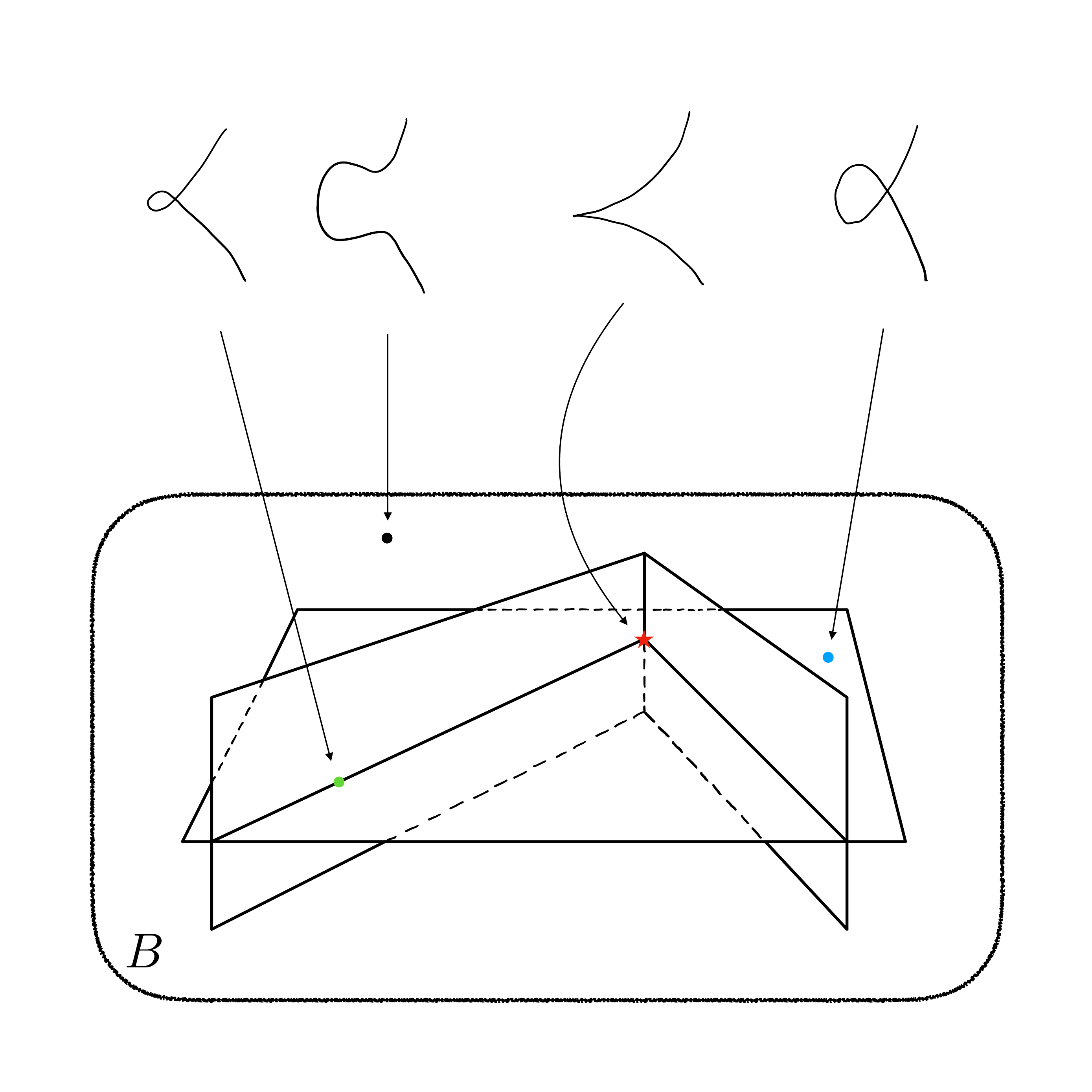}
	\caption{Elliptic fibration over a base $B$.
	While the fiber over the generic point (black dot) is smooth, it degenerates over codimension one (blue dot) loci, which corresponds to locations of 7-branes with a gauge symmetry.
	Intersections of 7-branes (green dot) form matter curves, where the fiber singularity enhances, indicating charged matter.
	Over codimension three points (red star), where matter curves intersect, further singularity enhancement signals Yukawa couplings.
	}\label{fig:full_fibration}
\end{figure}

Physically, the complex structure $\tau$ of every fiber $\pi^{-1}(p)$ specifies the value of the type IIB axio-dilaton $\tau = C_0 + \frac{i}{g_s}$ at $p$.
At codimension one subspace of the base, defined by the vanishing of the discriminant
\begin{align}\label{eq:discrimimant_weierstrass}
	\Delta := 4\,f^3 + 27\,g^2 \, ,
\end{align}
the elliptic fiber degenerates, signaling the presence of spacetime filling 7-branes which backreact onto $\tau$.
The resulting singularities encode the gauge dynamics of the 7-branes' world volume theory.
An enhancement of the singularity in codimension 2 signals the presence of matter states, while codimension 3 enhancements correspond to Yukawa couplings that are realized perturbatively in the effective field theory.
This set-up is summarized graphically in figure \ref{fig:full_fibration}.
Note that in the type IIB picture, the torus fiber is merely a bookkeeping device for the axio-dilaton.
However, through duality to M-theory, the torus actually becomes part of the physical compactification space.

\begin{figure}[!ht]
\centering
	\subfigure[$\mathfrak{su}(n-1)$ fiber]{\label{fig:A_n)fibre}
		\def\svgwidth{.44\hsize} 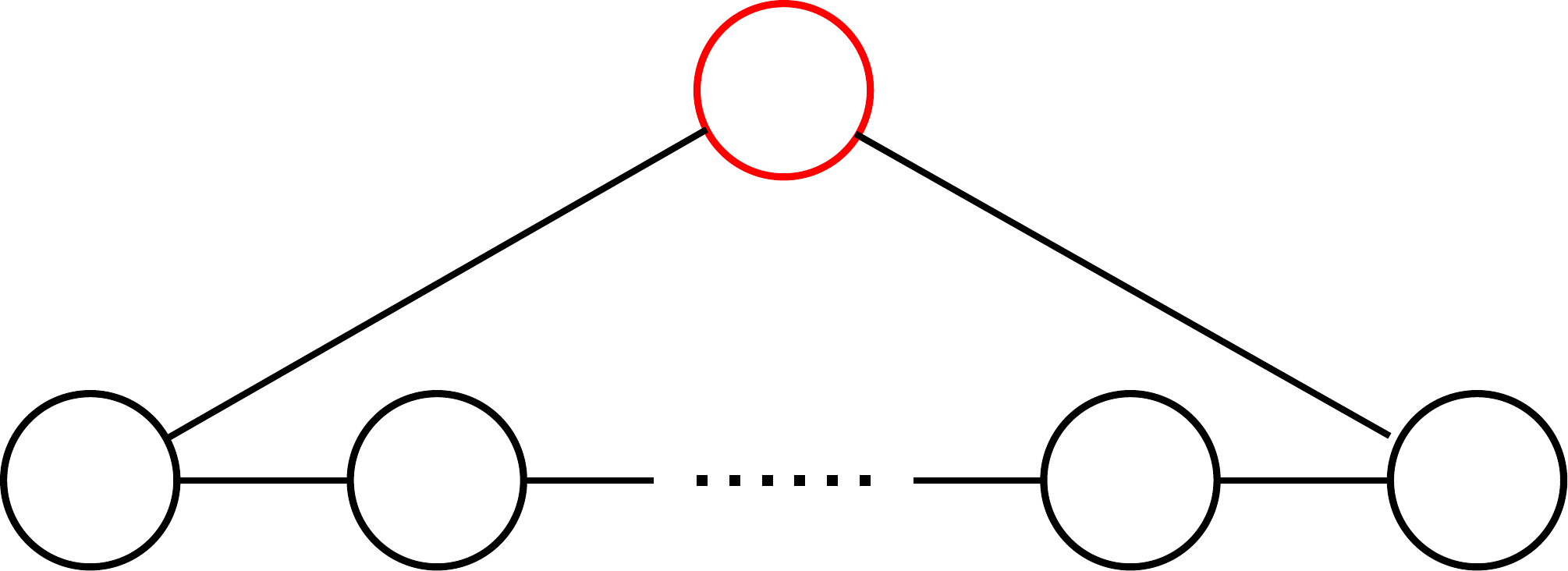 
	}
	\subfigure[$\mathfrak{sp}(n)$ fiber]{\label{fig:C_n)fibre}
		\def\svgwidth{.44\hsize} 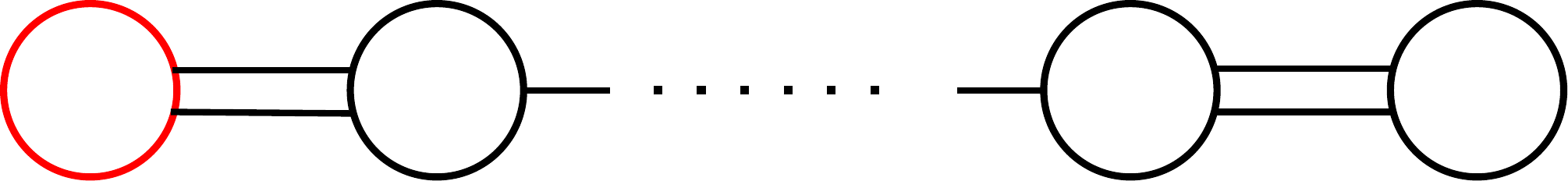 
	}
	\subfigure[$\mathfrak{so}(2n)$ fiber]{\label{fig:D_n)fibre}
		\def\svgwidth{.44\hsize} 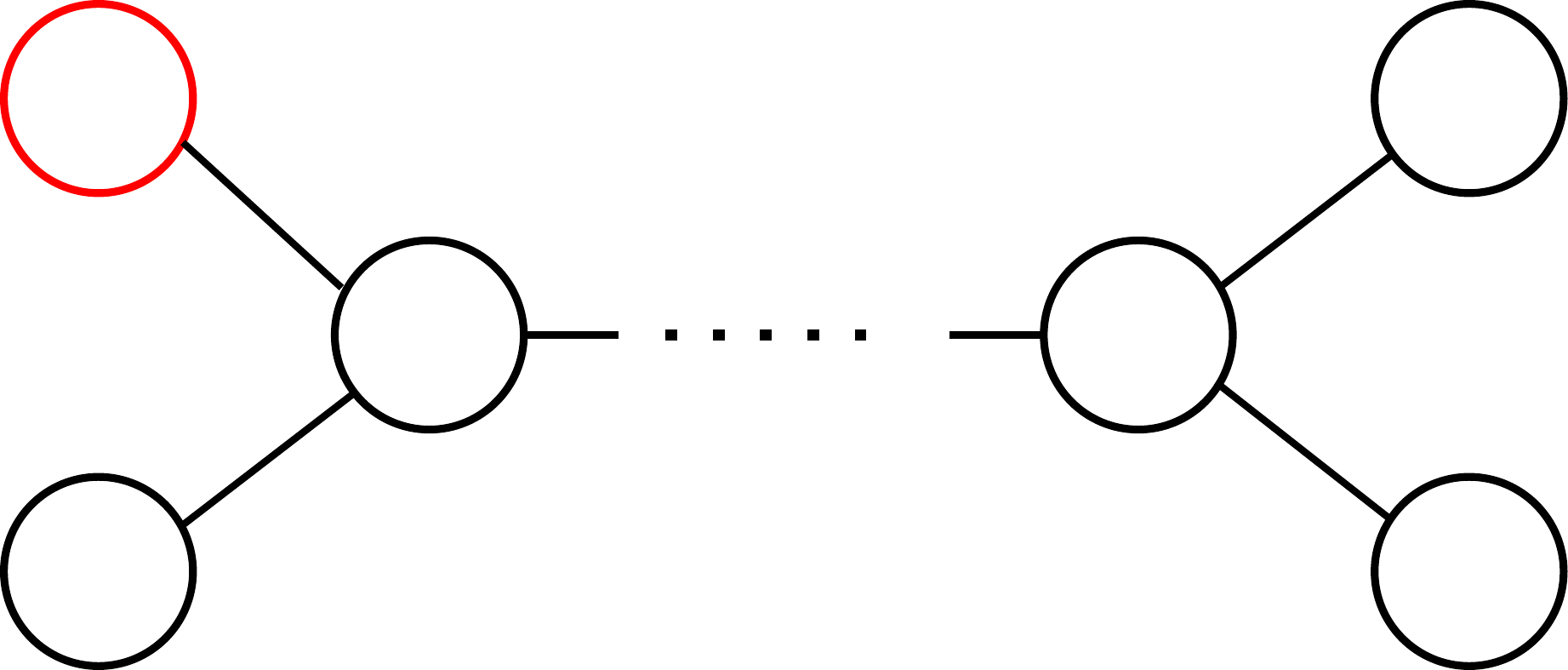 
	}
	\subfigure[$\mathfrak{so}(2n-1)$ fiber]{\label{fig:B_n)fibre}
		\def\svgwidth{.44\hsize} 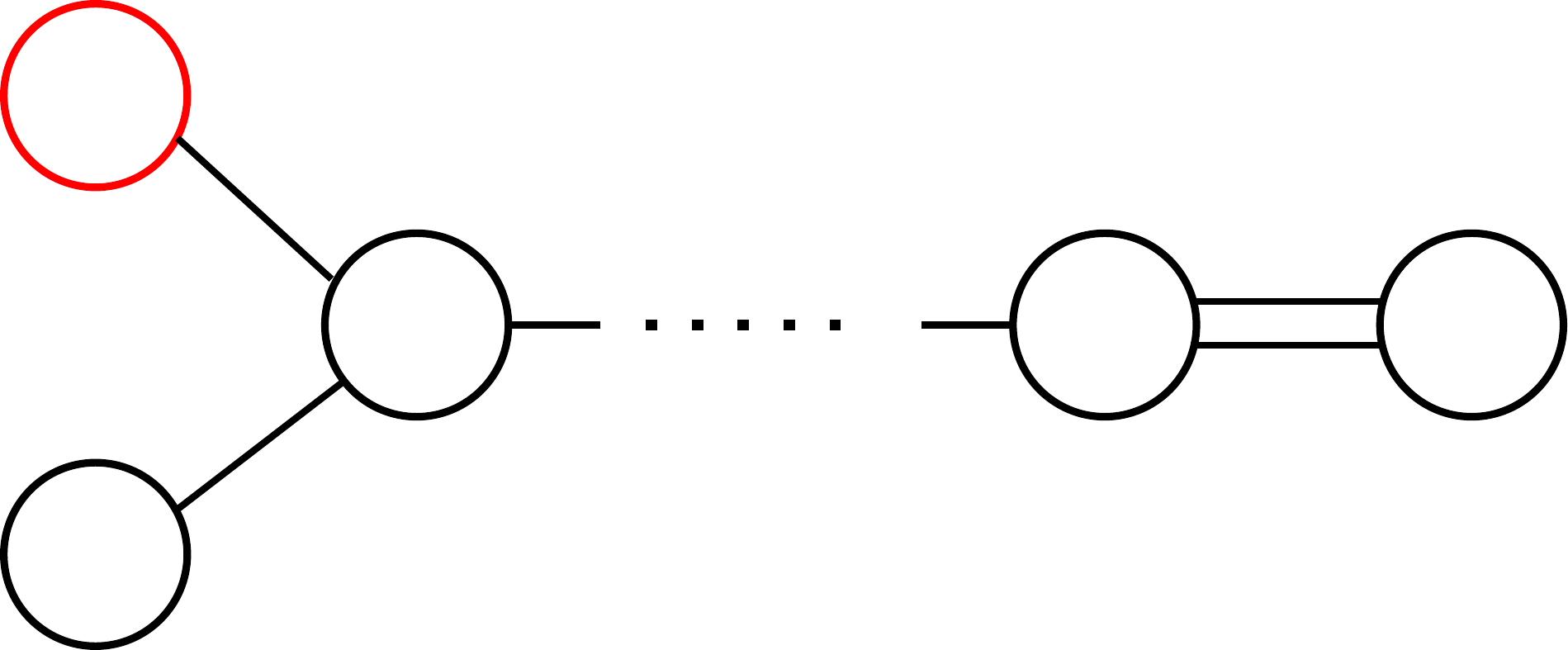 
	}
	\subfigure[$\mathfrak{e}_6$ fiber]{\label{fig:E_6)fibre}
		\def\svgwidth{.3\hsize} 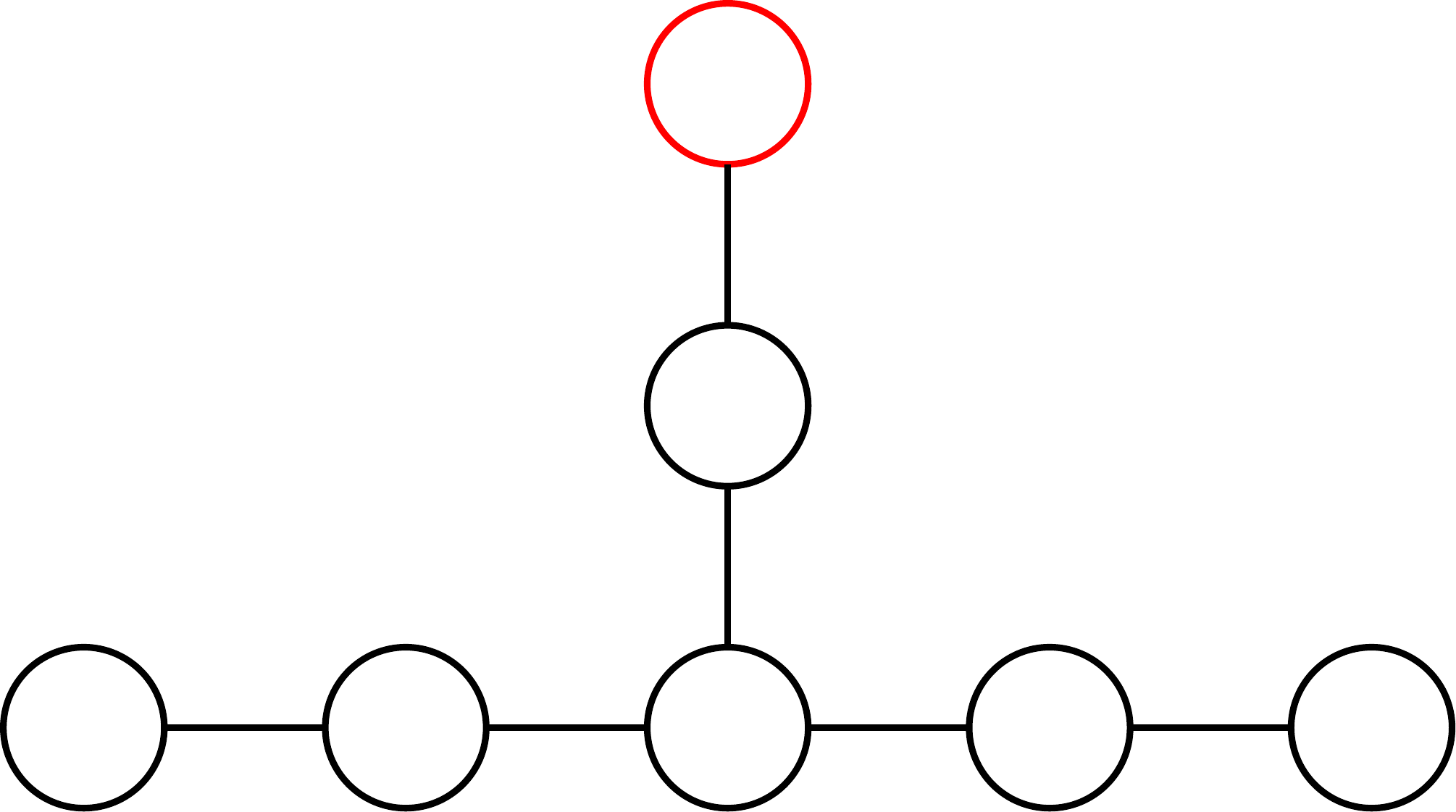 
	}
	\subfigure[$\mathfrak{e}_7$ fiber]{\label{fig:E_7)fibre}
		\def\svgwidth{.3\hsize} 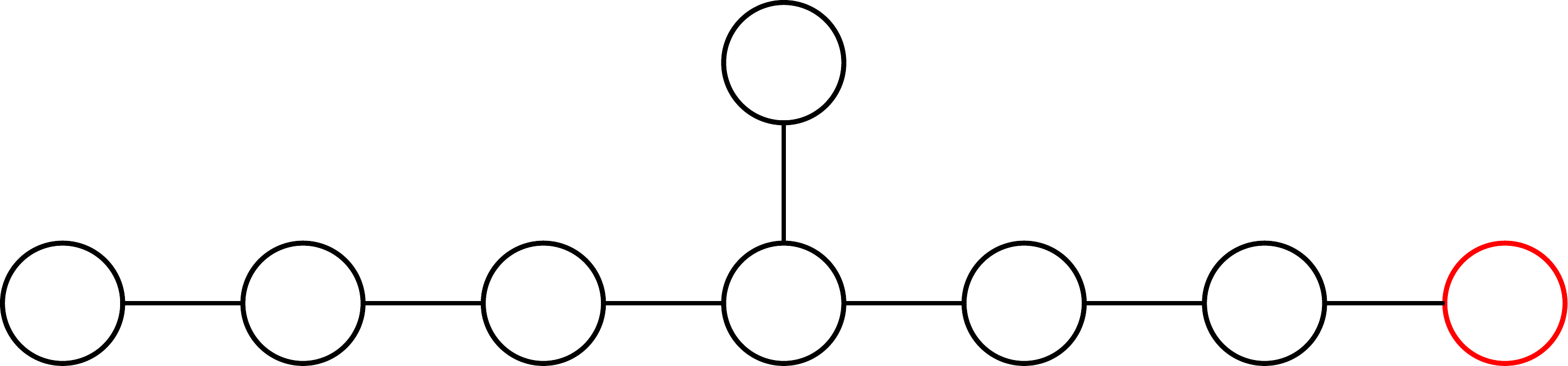 
	}
	\subfigure[$\mathfrak{e}_8$ fiber]{\label{fig:E_8)fibre}
		\def\svgwidth{.3\hsize} 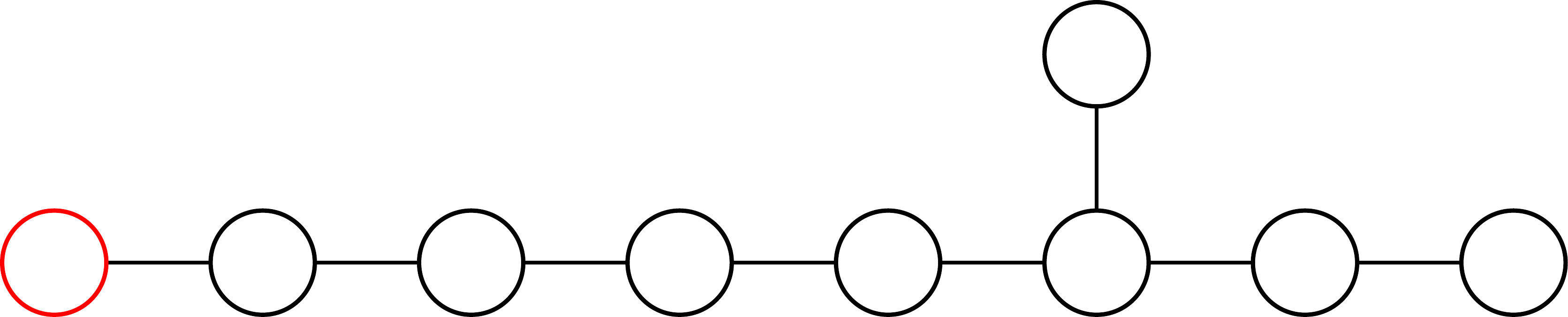 
	}
	\subfigure[$\mathfrak{f}_4$ fiber]{\label{fig:F_4)fibre}
		\def\svgwidth{.33\hsize} 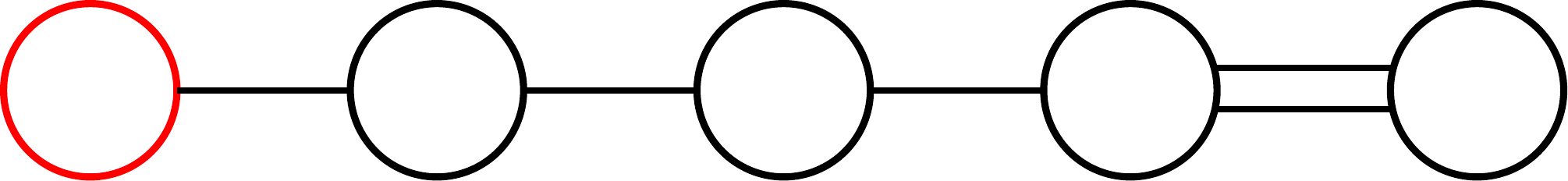 
	}
	\hspace{2cm}
	\subfigure[$\mathfrak{g}_2$ fiber]{\label{fig:G_2)fibre}
		\def\svgwidth{.2\hsize} 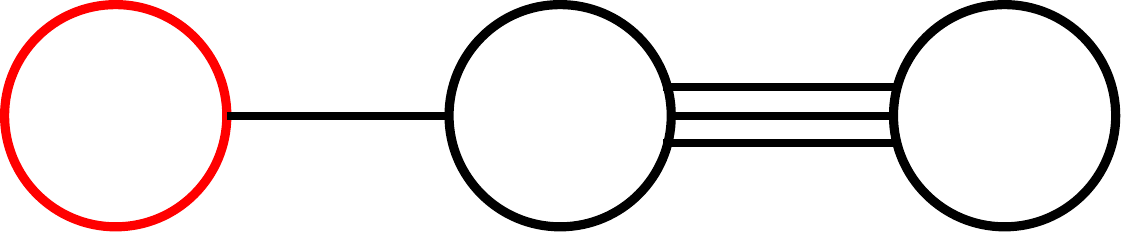 
	}
	\caption{
	Blow-up resolution of singular fibers take the form of the affine Dynkin diagrams of simple Lie algebras.
	Geometrically, each node represents a $\mathbb{P}^1$ component, with the multiplicity indicated by the number.
	Each line is a intersection point between the attached $\mathbb{P}^1$s; multiple lines correspond to higher intersection numbers. 
	The node in red marks the so-called affine node and is intersected by the zero section.
	This component of the fiber a pinched torus in the singular limit. 
	Note that for the diagrams \subref{fig:A_n)fibre} -- \subref{fig:B_n)fibre}, the number $n$ corresponds to the number of non-affine nodes.
	This is also the rank of the gauge group.}\label{fig:dynkin_diagrams}
\end{figure}

Concretely, the duality relates F-theory theory in $d = 12 - 2n$ dimension via a circle reduction to M-theory in $d-1$ \cite{Vafa:1996xn, Witten:1996bn}:
\begin{align}
	\text{F-theory on } \, Y_n \times S^1 \, \, \cong \, \, \text{M-theory on } \, Y_n \, .
\end{align}
A large part of the geometry/physics dictionary of F-theory can be best understood through this duality.
However, the interesting F-theory physics is encoded in the singularities of $Y_n$, which does not allow for a direct analysis in M-theory.
Instead, one first has to blow up the singularities of $Y_n$ to obtain a smooth space on which we can dimensionally reduce M-theory.
The blow up procedure introduces finite sized $\mathbb{P}^1$s at the singularities in the fiber over the discriminant locus $\{\Delta\}$.
Over an irreducible component $\Sigma$ of $\{\Delta\}$, the intersection pattern of these resolution $\mathbb{P}^1$s form the affine Dynkin diagram of an Lie algebra $\fkg_\Sigma$, see figure \ref{fig:dynkin_diagrams}.

Wrapping M2-branes on these $\bbP^1$s give rise to the W-bosons of the gauge symmetry, which after circle reduction are accompanied by a tower of massive Kaluza--Klein (KK) states.
These correspond to M2-branes which wrap, in addition to the $\bbP^1$s, the full torus fiber multiple times.
In codimension two, further singularities require small resolutions introducing additional $\bbP^1$s, on which wrapped M2-branes give rise to matter states in representations $\bf R$.
In the smooth phase of the geometry, these states as well as the W-bosons are massive.

Only in the singular limit, where the $\bbP^1$s all shrink to zero size, all W-bosons and matter states become massless.
While the fibral $\bbP^1$-curves introduced by the resolution account for (charged) W-bosons and matter states, the Cartan $\fku(1)$ gauge fields of $\fkg_\Sigma$ have a different origin.
By sweeping out each resolution $\bbP^1$ over the discriminant component $\Sigma$, we obtain $\text{rank}(\fkg_\Sigma)$ linearly independent divisors (complex codimension one subvarities) $E^{(\Sigma)}_i$ of $Y_n$.
Poincar\'{e}-duality implies that these divisors are in one-to-one correspondence to harmonic $(1,1)$-forms $\omega^{(\Sigma)}_i$.
Dimensionally reducing the M-theory 3-form, $C_3 = \sum_{\Sigma,i} \, \omega^{(\Sigma)}_i \wedge A_i^{(\Sigma)} + ...$, along these harmonic forms give rise to vector fields $A_i^{(\Sigma)}$ that uplift in the M-/F-duality to the Cartan gauge fields of $\fkg_\Sigma$ in the effective field theory of the F-theory compactification.

The geometry matches the representation theory in the following way.
For any holomorphic curve $\Gamma$, M2-branes wrapping these give rise to (in general massive) particle states carrying charges $c_i$ under the Cartan $\fku(1)$s given by
\begin{align}\label{eq:Cartan_charge_formula}
	c_i = \Gamma \cdot E^{(\Sigma)}_i \, .
\end{align}
If $\Gamma$ is one of the fibral $\bbP^1$s in codimension one, then $c_i$ form the weight vectors of the simple roots of $\fkg$, i.e., the ``charges vector'' of W-bosons under the Cartan $\fku(1)$s.
Fibral $\bbP^1$s localized in codimension two can have intersection numbers with $E^{(\Sigma)}_i$ which form weight vectors ${\bf w}$ of other representations $\bf R$.

\section[\texorpdfstring{\boldmath{$\mathfrak{u}(1)$}}{u(1)} Symmetries in F-theory]{\boldmath{$\mathfrak{u}(1)$} Symmetries in F-theory}\label{sec:u1}

As we have just seen, vector fields---the physical degrees of freedom of a gauge field---arise from dimensional reducing the M-theory $C_3$-form along harmonic $(1,1)$-forms $\omega$ dual to divisors $D$.
However, not all vector fields obtained this way remain massless when uplifting from M- to F-theory.
In fact, the masslessness condition require $\omega$ to have ``one leg along the base and one leg along the fiber'' \cite{Grimm:2010ez} of $Y_n$, which eliminates divisors $D = \pi^{-1}(D_B)$ pulled back from the base $B_{n-1}$ as sources of $\fku(1)$ symmetries.
Since the vectors associated with exceptional divisors are actually part of the full non-abelian gauge fields, the degrees of freedom of a genuine $\fku(1)$ symmetry has to come from somewhere else.

Indeed, there is a particular set of divisors that play a prominent role in the study of elliptic fibrations, namely so-called sections.
A section is a rational map $s: B_{n-1} \rightarrow Y_n$ from the base into the total space of the fibration, which marks one point in each fiber: $\pi \circ s = \text{id}_B$.
This defines a copy of the base $B_{n-1}$ inside of $Y_n$, and hence a divisor.
In fact, the Shioda--Tate--Wazir theorem \cite{MR2041769} states that in an elliptic fibration, up to linear equivalence the only divisors other than pull-backs and exceptionals are sections.
Explicitly, the rank of the N\'{e}ron--Severi group $\text{NS}(Y_n)$ --- the group of divisors modulo linear equivalence --- is given by
\begin{align}\label{eq:ShiodaTateWazir}
	\text{rk(NS}(Y)) = \underbrace{\text{rk(NS}(B))}_{\text{pull-back}} + \underbrace{\sum_S \text{rank}(\fkg_S)}_{\text{exceptional}} + \underbrace{1 + \text{rk(MW}(Y))}_{\text{sections}} \, ,
\end{align}
where we have used F-theory language to count the number of independent exceptional divisors by the rank of the non-abelian gauge algebra.
Note that the notation already indicates that the sections form an abelian group ``MW'' which has finite rank.
The structure of this so-called Mordell--Weil group plays a central role in the discussion of $\fku(1)$ symmetries in F-theory.

\subsection{The Mordell--Weil group of rational sections}

The most intuitive way to see that sections form an abelian group is to map the elliptic fiber $\mathfrak{f}_t$ to a torus $T_t^2 \cong \mathbb{C}/\Lambda_t$, where $\Lambda_t$ is a two (real) dimensional lattice.\footnote{These and other well-known properties of elliptic curves and fibrations can be found in standard text books, e.g., \cite{MR2514094, MR1312368}.}
Under this map, sections map fiberwise to points on the fundamental domain of the torus $T^2_t$, which is just a patch of $\mathbb{C}$.
For points in $\mathbb{C}$, there is a natural abelian group law given by simple addition.
By mapping the result of the addition back to the elliptic fiber, one obtains another section.

In this picture, we have implicitly agreed on a common zero element on each fiber $\mathfrak{f}_t$, which maps onto the origin of the quotient $\mathbb{C}/\Lambda_t$ for any $t \in B$.\footnote{More precisely, for any $t$ up to codimension two loci of $B$.}
This common zero element is itself a section, usually referred to as the zero section.
One is in principle free to choose the zero section, which does not change the arithmetic structure of the Mordell--Weil group.
However, as we will discuss later, there is non-trivial physical information associated with this freedom to choose the zero section.
In any case, with the choice of a zero section, one can find the corresponding Weierstrass model \eqref{eq:weierstrass_equation}, with the chosen zero section mapped to $\sigma_0 : [x:y:z] = [1:1:0]$.

\begin{figure}[ht]
	\centering
	\def\svgwidth{.4\vsize}
	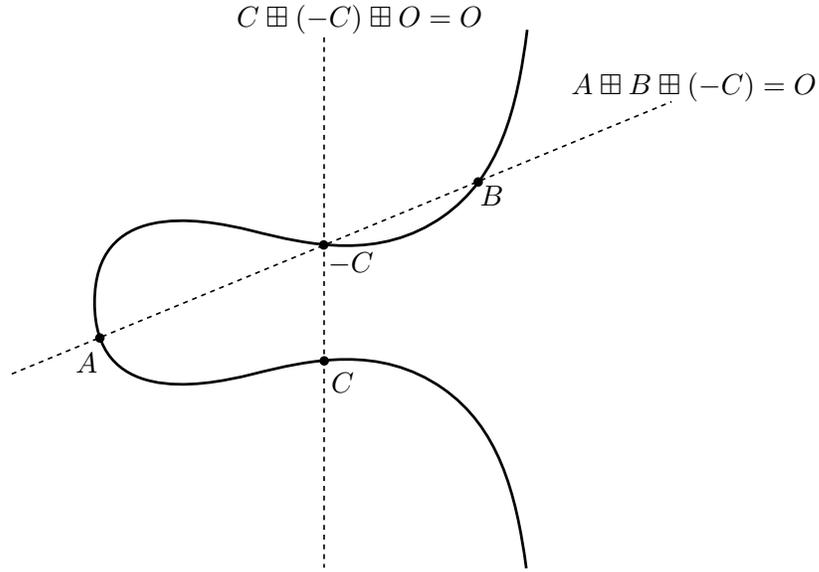
	\caption{Geometric construction of the Mordell--Weil group law. Each dashed line marks three points on the elliptic curve (solid curve) that add up to zero under the group law. The rational points $A,B,C$ satisfy $A \boxplus B = C$.}
	\label{fig:elliptic_group_law}
\end{figure}

In the Weierstrass form, there is another geometric way of defining the group law of sections.
To do so, we again look at each fiber individually.
In the $z=1$ patch of $\bbP_{231}$, the point marked by the zero section is the point $O$ at infinity.
The group law $\boxplus$ is defined by declaring that three points $A, B, C \in E$, which also lie on a straight line in the $x$-$y$-plane, satisfy $A \boxplus B \boxplus C = O$.
To add up two points, one has to take into account that a vertical line will meet $E$ at infinity, i.e., $O$.
This geometric realization of the group law is depicted in figure \ref{fig:elliptic_group_law}.
It is straightforward to check that $\boxplus$ defined this way satisfies all properties (associativity, commutativity, unique inverse element) necessary for an abelian group.

The above fiberwise construction can be extended across the whole base $B$ of the elliptic fibration $Y$.
However, not every point on a fiber $\mathfrak{f}_t$ can be the image $s(t)$ of a section $s: B \rightarrow Y$.
Because $s$ has to be a rational map, the Weierstrass coordinates $[x:y:z]$ of $s(t)$ must be meromorphic functions on $B$.
The arithmetic description of elliptic fibrations explains the attribute ``rational'' more clearly.
Namely, an elliptic fibration over $B$ can be also viewed as an elliptic curve over the function field $K(B)$ of the base.
Elements $q \in K(B)$ are called rational functions, because on any open patch of $B$ they can be written as quotients $q = \frac{p_1}{p_2}$ of global sections of some line bundles; in a local chart, the $p_i$s can be written as polynomials in the local coordinates.
A section of the elliptic fibration is then a \textit{rational} solution of the Weierstrass equation \eqref{eq:weierstrass_equation}, meaning there are $x_Q, y_Q, z_Q \in K(B)$ such that $y_Q^2 = x_Q^3 + f\,x_Q\,z_Q^4 + g\,z_Q^6$.

The abelian group constructed this way is called the Mordell--Weil group $\text{MW}(Y)$ of the elliptic fibration $Y$.
By the famous Mordell--Weil theorem, this group is finitely generated:
\begin{align}
	\text{MW}(Y) = \bbZ^{\oplus r} \oplus \bbZ_{k_1} ... \oplus \bbZ_{k_t} \, .
\end{align}
The rank of the Mordell--Weil group is the number $r$ of independent \textit{free} generators.
By the Shioda--Tate--Wazir theorem \eqref{eq:ShiodaTateWazir} these are the only independent divisors in addition to the exceptional and pull-back divisors.
They are to be distinguished from torsional generators $\tau_{k_i}$, for which there is a (minimal) positive integer $k_i$ such that $\sigma_0 = \tau_{k_i} \boxplus ... \boxplus \tau_{k_i}$ ($k_i$ times), where $\sigma_0$ is the zero section.
The divisor classes of these sections are linearly dependent with other divisors, and we will come back to the physical implication of this fact in a moment.
Note that in \eqref{eq:ShiodaTateWazir}, the contribution of sections to the N\'{e}ron--Severi rank was $1+r$.
This is due to the nature of the zero section, which is an independent section, but---as it is the neutral group element---does not contribute to the rank of the Mordell--Weil group.

\subsubsection{Example: The U(1)-restricted Tate model}\label{sec:example_restricted_Tate}

Before we move on, let us look at a simple example from the F-theory literature of an elliptic fibration with non-trivial Mordell--Weil group.
This so-called $U(1)$-restricted Tate model was first introduced in \cite{Grimm:2010ez} and given by the equation
\begin{align}\label{eq:U1-restricted_tate}
	y^2 + a_1\,x\,y\,z + a_3\,y\,z^3 = x^3 + a_2\,x^2\,z^2 + a_4\,x\,z^4 \, ,
\end{align}
where $[x:y:z]$ are homogenous coordinates of $\bbP_{231}$, and $a_i$ are sections of the line bundles $K_B^{\otimes (-i)}$.
In addition to the zero section $[x:y:z] = [1:1:0]$, there is now also an additional rational section at $[x:y:z] = [0:0:1]$.
Note that the equation \eqref{eq:U1-restricted_tate} is not in Weierstrass form!
For that, one has to perform a birational transformation, which also shifts the coordinates of the fiber ambient space.
The resulting Weierstrass functions are
\begin{align}\label{eq:U1-restricted_weierstrass}
	\begin{split}
		f & = \frac{a_1\,a_3}{2} + a_4 - \frac{1}{48} \, (a_1^2 + 4\,a_2)^2 \, , \\
		g & = \frac{1}{864} \left( (a_1^2+4\,a_2)^3 + 216\,a_3^2 - 36\,(a_1^2 + 4\,a_2)\,(a_1\,a_3 + 2\,a_4) \right) \, .
	\end{split}
\end{align}
The corresponding Weierstrass equation \eqref{eq:weierstrass_equation} then has the rational solution
\begin{align}\label{eq:weierstrass_coord_U1-restricted}
	[x_Q : y_Q : z_Q] = \left[ \frac{a_1^2 + 4\,a_2}{12} : \frac{a_3}{2} : 1 \right] \, .
\end{align}

\begin{figure}[ht]
	\centering
	\def\svgwidth{.6\vsize}
	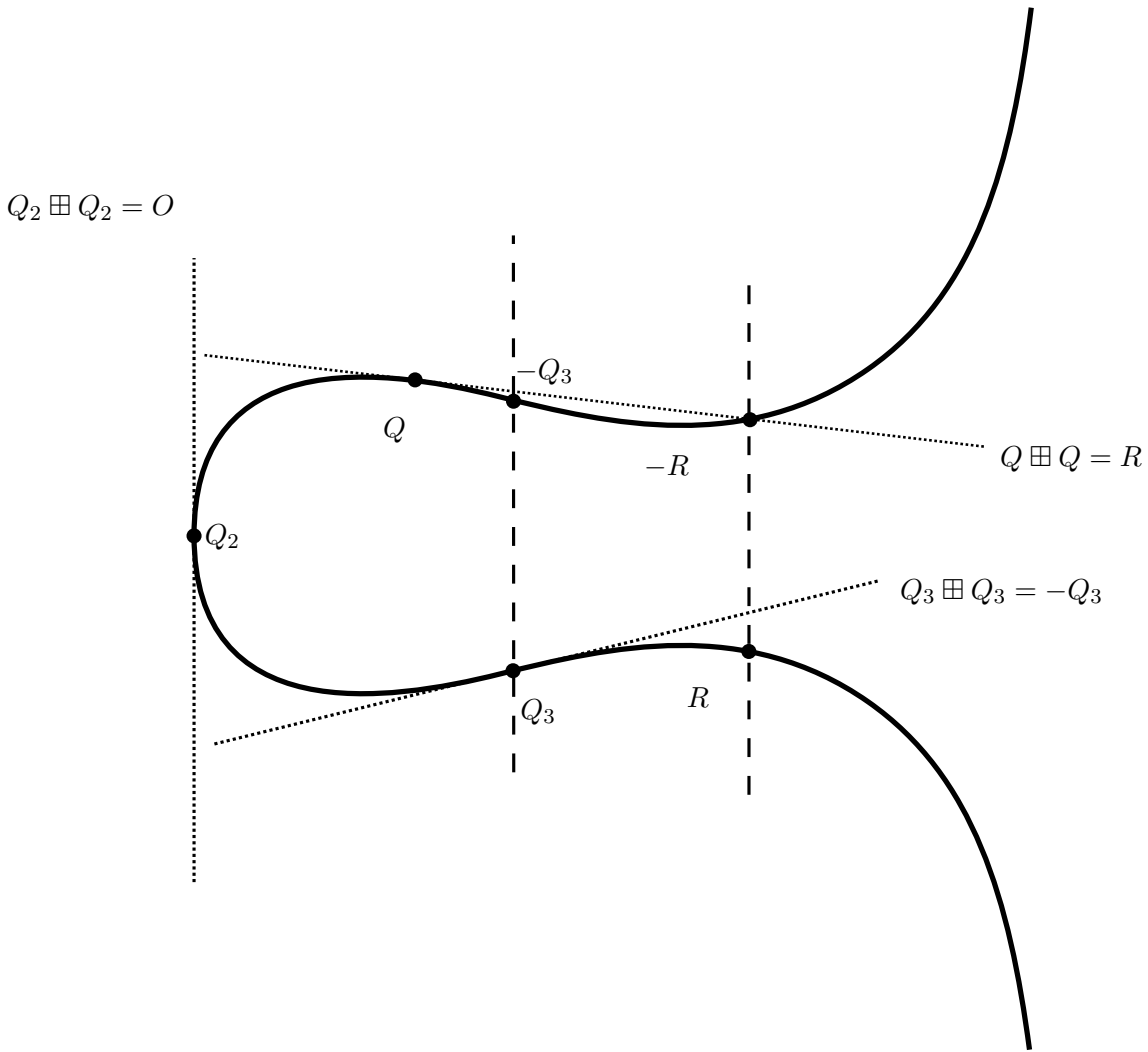
	\caption{A 2-torsional point $Q_2$ on an elliptic curve has to have a vertical tangent. A 3-torsional point $Q_3$ is a point of inflection.}
	\label{fig:MW-torsion}
\end{figure}

For generic choices of coefficients $a_i$, this rational section generates the Mordell--Weil group $\mathbb{Z}$.
However, we can tune the model such that the sections becomes 2-torsional, i.e., the Mordell--Weil group is $\bbZ_2$.
This is achieved by setting $a_3 \equiv 0$ globally.
How do we see that this turns the section into an element of order two?
To answer that, first observe that with this tuning, the $y$-coordinate of the section \eqref{eq:weierstrass_coord_U1-restricted} becomes 0 everywhere.
This means that in every fiber (up to higher codimension), the rational point has a vertical tangent in the $x$-$y$-plane, because the (smooth) cubic $y^2 = x^3 + f\,x + g$ has infinite slope at $y=0$.\footnote{Taking the total derivative in the $x$-$y$-plane for the Weierstrass equation yields $2\,y\,\text{d}y = (3\,x^2 + f)\,\text{d}x$.
Because the elliptic curve is smooth by assumption (it is the generic fiber), $y$ and $3\,x^2 + f$ cannot vanish simultaneously.
This means however, that $\text{d}y/\text{d}x = (3\,x^2 + f)/2y$ diverges at $y=0$.}
However, a vertical tangent at the point $Q$ precisely means $Q \boxplus Q \boxplus O = O \Leftrightarrow Q\boxplus Q = O$ under the group law, cf.~figure \ref{fig:MW-torsion}, implying that the section an element of order two in the Mordell--Weil group.
Likewise, one could also imagine tuning the rational section to sit at a point of inflection on the generic fiber, which under the group law constitutes an element of order three.
Thus, the Mordell--Weil group in this case would be $\bbZ_3$.

In physical terms, this kind of complex structure deformation correspond to a gauge enhancement (sometimes also called unHiggsing) of the $\fku(1)$ into a non-abelian algebra.
To understand this statement, we first have to discuss how exactly the information contained by the Mordell--Weil group is mapped into physical data about gauge symmetries.

\subsection{The Shioda map}

Recall that by the Shioda--Tate--Wazir theorem \eqref{eq:ShiodaTateWazir}, the number of independent divisors that do not arise from exceptional or pull-back divisors is $1 + \text{rk(MW)}$.
The divisor class $Z$ of the zero section is dual to the Kaluza--Klein $\fku(1)$ that arises in the circle compactification of the F-/M-theory duality.
Heuristically, one can then identify a (free) MW-generator $\sigma$ as the dual divisors of $\fku(1)$ gauge symmetries in F-theory.
However, to properly specify the massless vector field which furnishes these $\fku(1)$s, the dual divisor class $\varphi(\sigma)$ has to satisfy the following consistency conditions:
\begin{align}\label{eq:transversality_U1_generator}
	\begin{split}
		(1) & \quad \varphi(\sigma) \cdot \mathfrak{f} = 0 \, ,\\
		(2) & \quad \varphi(\sigma) \cdot {\cal C}_B = 0 \, ,\\
		(3) & \quad \varphi(\sigma) \cdot \bbP^1_i = 0 \, .
	\end{split}
\end{align}
The first condition, imposing vanishing intersection number of $\varphi(\sigma)$ with the generic fiber $\mathfrak{f}$, ensures that all Kaluza--Klein tower states in M-theory that originate from the same states in F-theory have the same $\fku(1)$ charge under $\varphi(\sigma)$.
The second condition, imposing vanishing intersection number with any curve ${\cal C}_B$ in the base, ensures that there are no axionic gaugings of the $\fku(1)$ which would lead to a mass term.
The first two conditions are the mathematical description of $\varphi(\sigma)$ having ``one leg along the fiber and one along the base'' \cite{Grimm:2010ez}.
Finally, the third condition, imposing vanishing intersection with the fibers of exceptional divisors, ensures that no W-boson of the non-abelian gauge symmetries is charged under the $\fku(1)$.
These conditions are a consequence of the general formula for $\fku(1)$ charges of matter states coming from M2-branes wrapping a holomorphic curve $\Gamma$, which similar to the case of Cartan $\fku(1)$s \eqref{eq:Cartan_charge_formula} now reads
\begin{align}\label{eq:U1-charge-general}
	q = \Gamma \cdot \varphi(\sigma) \, .
\end{align}

Given a section $\sigma$, these three conditions determine $\varphi(\sigma)$ up to an overall normalization.
Remarkably, the same conditions have been considered in the mathematics literature \cite{MR1030197, MR1081832}, which leads to the so-called Shioda map.
This map associates a unique divisor class $\varphi(\sigma)$ to a section $\sigma$ compatible with the Mordell--Weil group law (i.e., it is a group homomorphism $\text{MW}(Y) \stackrel{\varphi}{\longrightarrow} \text{NS}(Y)$):
\begin{align}\label{eq:shioda-homomorphism}
	\varphi(\sigma_1 \boxplus \sigma_2) = \varphi(\sigma_1) + \varphi(\sigma_2) \, .
\end{align}
One can fix the normalization by requiring $\varphi(\sigma) = S + ...$, where $S=[\sigma]$ is the divisor class of the section.
Then the map takes the form
\begin{align}\label{eq:Shioda-map}
	\varphi(\sigma) = S - Z  - \pi( (S - Z) \cdot Z) + \sum_k \lambda_k \, E_k \, .
\end{align}
Here, the term $\pi( (S-Z) \cdot Z)$ is the projection of the 4-cycle class $[(S-Z) \cap Z]$ to a divisor on the base $B$, and guarantees condition (2) in \eqref{eq:transversality_U1_generator}.
Its explicit form depends on the geometry, but for the purpose of these notes, it suffices to say that this term is a divisor pulled-back from the base, which does not intersect any fibral curves, hence does not contribute to the charges of states.\footnote{
However, the volume of the divisor in the base encodes information about the gauge coupling of the $\fku(1)$, and is important in the recent geometric proof that $\fku(1)$ symmetries cannot be strongly coupled in 6D \cite{Lee:2018ihr}.}

In the following, we will focus on the term $\lambda_k \, E_k$, which has some interesting physical implications.
Recall that the exceptional divisors $E_k$ are $\bbP^1$ fibrations over a codimension one locus $W \subset B$.
Wrapping the fiber component $\bbP^1_k$ of $E_k$ with M2-branes gives rise to the gauge bosons of the non-abelian gauge algebra $\fkg$ over $W$.
As they carry weights of the simple roots $-\alpha_k$ of $\fkg$, their intersection matrix $E_i \cdot \bbP^1_j = -C_{ij}$ is the negative Cartan matrix of $\fkg$.\footnote{
If $\fkg = \bigoplus_l \fkg_l$ is a sum of simple algebras, then the Cartan matrix is the block-diagonal matrix formed by the Cartan matrices of $\fkg_l$.
}
In order to ensure that the gauge bosons of $\fkg$ are not charged under the $U(1)$, i.e., to satisfy condition (3) in \eqref{eq:transversality_U1_generator}, the coefficients $\lambda_k$ can be explicitly determined to be
\begin{align}\label{eq:fractional_coefficients_shioda}
	\lambda_k = \sum_l \left( (S-Z) \cdot \bbP_l \right) \, (C^{-1})_{lk} \, .
\end{align}
These coefficients depend on the different intersection structure between the sections $S$ and $Z$ with the fiber components of the exceptional divisors.
In general, they will be fractional numbers, since it involves the inverse Cartan matrix $C^{-1}$.
As a consequence, $\lambda_k \in  \frac{1}{N} \bbZ$ for all $k$, where $N$ depends on $\fkg$ and the ``fiber split type'' \cite{Braun:2013nqa} given by the numbers $(S-Z) \cdot \bbP^1_l$.\footnote{It is called ``fiber split'', because these numbers encodes how the section $\sigma$ intersects the codimension one fiber $\bbP^1$s differently than the zero section.}
But importantly, it is always finite and can be chosen to be minimal, i.e., the numerators of all $\lambda_k$ have greatest common divisor 1.

\subsubsection{The Shioda map as a lattice embedding}\label{sec:lattice_embedding}

In this short section, we briefly review the original mathematical work \cite{MR1030197, MR1081832} that motivated the Shioda map.
As the details are not immediately relevant for the rest of the notes, it can be safely skipped.

The original motivation of Shioda to introduce the map \eqref{eq:Shioda-map} was to identify the Mordell--Weil group as a ``sublattice'' of the N\'{e}ron--Severi group.
More precisely, in the arithmetic description of elliptic curves, there is a so-called height pairing (see, e.g., \cite{MR2514094}) defined on the Mordell--Weil group,
\begin{align}
	\langle \cdot , \cdot \rangle: \text{MW} \times \text{MW} \longrightarrow \mathbb{R} \, ,
\end{align}
which induces a lattice structure on $\text{MW}/\text{Tors(MW)}$, where Tors(MW) denotes the torsion part of Mordell--Weil.

On the other hand, for an elliptic surface, there is also a natural ``algebraic'' pairing of sections given by the intersection product, which defines the lattice structure on the N\'{e}ron--Severi group.
Shioda showed that the two different pairings can be identified, by embedding the Mordell--Weil group into the N\'{e}ron--Severi lattice.
However, the embedding cannot be injective, because the Mordell--Weil group has torsion whereas the N\'{e}ron--Severi group does not.
This is remedied by considering the quotient $\text{NS}/T$, where $T$ is generated by the zero section $Z$, all pull-back divisors $D_B$ and all exceptional divisors $E_i$.
Note that these are precisely the divisors dual to the curves which must have intersection number 0 with the Shioda map \eqref{eq:transversality_U1_generator}!

With this sublattice $T$, Shioda proved the isomorphism
\begin{align}
	\text{MW}(Y) \cong \text{NS}(Y) / T \, ,
\end{align}
For the proof, he introduced the map $\varphi$ to ``split'' this isomorphism: 
\begin{align}
	\text{NS}(Y) = \text{Im}(\varphi) \oplus_{\perp} T \, ,
\end{align}
where $\oplus_{\perp}$ indicates that the two summands are orthogonal with respect to the intersection pairing.
Because $\varphi(\text{Tors(MW)}) = 0$, it identifies, as promised, a sublattice of $\text{NS}(Y)$ with $\text{Im}(\varphi) = \text{MW}/\text{Tors(MW)}$.

Crucially, the map \eqref{eq:Shioda-map}---with the normalization set to 1---satisfies the identity
\begin{align}\label{eq:lattice_identification_shioda}
	\langle \sigma_1, \sigma_2 \rangle = - \varphi(\sigma_1) \cdot \varphi(\sigma_2) \, .
\end{align}
In other words, the arithmetic pairing $\langle \cdot , \cdot \rangle$ defines the same lattice on the Mordell--Weil group as the algebraic (intersection) pairing on Im($\varphi$).
Clearly, this identification would be spoiled by a rescaling of the Shioda map \eqref{eq:Shioda-map}.

The same identification can be generalized to higher dimensions.
However, the height pairing must now be modified to map onto the divisor group of the base $B$ of the fibration \cite{MR2041769}.
Likewise, the intersection product $\varphi(\sigma_1) \cdot \varphi(\sigma_2)$ is now a 4-cycle, which can also be pushed-down onto the base to give rise to a divisor.
Then, one can again identify the two resulting pairings via the Shioda map with normalization 1.

As we will see now, this lattice structure of the Mordell--Weil group manifest itself in the physics of F-theory compactifications in terms of the global gauge group structure.
%

\subsection{The global gauge group of F-theory}

So far, we have only mentioned the gauge algebra of the F-theory compactification.
The reason is that in general, the gauge group need not to be the naive simply connected Lie group associated with the algebra.
Rather, it takes the form
\begin{align}\label{eq:global_gauge_group_summary}
	\frac{U(1)^r \times G}{\prod_{i=1}^r \bbZ_{m_i} \times \prod_{j=1}^{t} \bbZ_{k_j} } \, .
\end{align}
This notation means that each discrete $\bbZ_n$ factor is a subgroup of $U(1)^r \times G$ which acts trivially on any matter representation.
In F-theory, the information about the global structure of the gauge group is encoded in the Shioda-map \eqref{eq:Shioda-map}, or more precisely, in the coefficients $\lambda_i$ \eqref{eq:fractional_coefficients_shioda} \cite{Mayrhofer:2014opa, Cvetic:2017epq}.
In anticipation of the result, we have already separated in \eqref{eq:global_gauge_group_summary} the contributions $\bbZ_{m_i}$ of the free Mordell--Weil generators from those of torsional generators $\bbZ_{k_j}$.
Because each factor is independent of the others, we will restrict our discussion below to cases with a single generating section, and refer to \cite{Cvetic:2017epq} for examples with multiple Mordell--Weil generators.

First, recall that matter states in F-theory arise from M2-branes wrapping curve components $\Gamma$ of reducible fibers in codimension two.
Since these curves are integral in homology (they are irreducible holomorphic subvarieties of the total space), their intersection numbers with any integral divisor, in particular the exceptional ``Cartan'' divisors and the sections, must be integral as well.
This implies that the intersection number of $\Gamma$ with the Shioda divisor \eqref{eq:Shioda-map} must satisfy
\begin{align}\label{eq:integrality_condition_shioda_map}
	\varphi(\sigma) \cdot \Gamma - \sum_k \lambda_k \, E_k \cdot \Gamma = (S - Z) \cdot \Gamma \in \bbZ \, .
\end{align}
Recall that $\Gamma$ corresponds to a weight ${\bf w}$ of a representation ${\bf R}$ of the non-abelian gauge algebra $\fkg$, which in the Dynkin basis is a vector with entries ${\bf w}_k = E_k \cdot \Gamma$, $k=1,... \text{rank} (\fkg)$.
Therefore, the condition \eqref{eq:integrality_condition_shioda_map} relates the non-abelian representation of $\Gamma$ with the intersection number $\varphi(\sigma) \cdot \Gamma$.

To see the physical relevance of this condition, we differentiate between the cases where the section $\sigma$ is generator of the torsional or the free part of the Mordell--Weil group.
In case $\sigma$ is $n$-torsional, the homomorphism property \eqref{eq:shioda-homomorphism} implies
\begin{align}
	n \, \varphi(\sigma) = \varphi( \underbrace{\sigma \boxplus ... \boxplus \sigma}_{ \times n}) = \varphi(0) = 0 \, .
\end{align}
But because the divisor group is torsion free, we must have $\varphi(\sigma)=0$.
In this case, the integrality condition \eqref{eq:integrality_condition_shioda_map} simply becomes:
\begin{align}\label{eq:integrality_condition_torsion}
	\sigma \, \text{ torsional}: \quad \sum_k \lambda_k \, {\bf w}_k \in \bbZ \, .
\end{align}
If $\sigma$ is a generator of the free part of Mordell--Weil, then the divisor $\varphi(\sigma)$ is dual to the massless $\fku(1)$ gauge field, and $\varphi(\sigma) \cdot \Gamma$ is the charge of the state on $\Gamma$.
Thus, the condition now becomes
\begin{align}\label{eq:integrality_condition_free}
	\sigma \, \text{ free}: \quad q_\sigma - \sum_k \lambda_k \, {\bf w}_k \in \bbZ \, .
\end{align}
The significance of these two conditions, which have also been noted in \cite{Grimm:2015wda} in a different context, are hidden in the coefficients $\lambda_k$.
As we will see now, these coefficients are related to the center of the non-abelian gauge symmetry.
More precisely, they define an element in the center ${\cal Z}(G)$, where $G$ is the simply connected Lie group with algebra $\fkg$.

\subsubsection{Center of the non-abelian gauge symmetry}

The crucial property of the $\lambda_k$ for constructing the center is the non-integral part of the sum $\sum_k \lambda_k \, {\bf w}_k$, which is the same for \text{any} state in the same representation ${\bf R}$ of $\fkg$.
In other words, we can define a fractional number between 0 and 1 via
\begin{align}\label{eq:capital_L_definition}
	L({\bf R}) = \sum_k \lambda_k \, {\bf w}_k \, \text{ mod } \bbZ \, , \quad {\bf w} \in {\bf R} \, ,
\end{align}
which is independent of the choice ${\bf w}$ and only depends on the representation $\bf R$ of $\fkg$.
To see this, we use the basic fact that two weights ${\bf w}, {\bf v} \in {\bf R}$ differ by an integer linear combination $\mu_i \, \alpha_i$ of the simple roots $\alpha_i$ of $\fkg$.
Geometrically, this means that the two fibral curves $\Gamma_{\bf w}$, $\Gamma_{\bf v}$ differ by a linear combination of the codimension one fibral $\bbP^1$s:
\begin{align}
	{\bf v} = {\bf w} + \sum_i \mu_i\,\alpha_i \, \Leftrightarrow \, \Gamma_{\bf v} = \Gamma_{\bf w} + \sum_i \mu_i \, \bbP^1_i \, , \quad \mu_i \ \in \bbZ \, .
\end{align}
Plugging in the explicit formula \eqref{eq:fractional_coefficients_shioda} for $\lambda_k$ as well as the relationship $E_k \cdot \bbP_i = -C_{ki}$, we obtain
\begin{align}
\begin{split}
	\sum_k \lambda_k \, {\bf v}_k & \, = \sum_k \left( \vphantom{\sum_l} \right. \sum_l ( \underbrace{(S - Z)\cdot \bbP_l}_{=:t_l \in \bbZ} ) (C^{-1})_{lk} \left. \vphantom{\sum_k} \right) \, E_k \cdot \Gamma_{\bf v} \\
	& = \sum_k \sum_l t_l \, (C^{-1})_{lk} \left( E_k \cdot \Gamma_{\bf w} - \sum_i \mu_i \, C_{ki} \right) \\
	& = \sum_k \lambda_k \, {\bf w}_k + \underbrace{\sum_l t_l \, \mu_l}_{\in \bbZ} \, ,
\end{split}
\end{align}
which shows that \eqref{eq:capital_L_definition} is well-defined.

One essential feature of the fractional number $L({\bf R})$ is that $N \times L({\bf R}) \in \bbZ$ for any representation ${\bf R}$.
The integer $N$ arises from taking the inverse Cartan matrix for defining $\lambda_k$, and depends on the fiber split type.
For example, if $\fkg = \mathfrak{su}(m)$, then $N$ is divisor of $m$; for $\fkg = \mathfrak{e}_{6/7/8}$, $N$ is 1 or $3/2/1$, respectively

Having established that, we can now construct an element of the center of $G$, the unique simply connected Lie group with algebra $\fkg$.
To do so, we define its action in each representation ${\bf R}$ of $\fkg$ via
\begin{align}\label{eq:central_element_non-ab_definition}
	{\bf w} \mapsto C \, {\bf w} := \left[ \exp(2\,\pi\,i\,L({\bf R}) ) \times  {\mathbb{1}_{\bf R}} \right] {\bf w} \, ,
\end{align}
where ${\bf w}$ is any weight of ${\bf R}$, i.e., any basis vector of the representation space of ${\bf R}$.
It can be shown that this action really is the exponentiation of a linear action of a Lie algebra element in the representation ${\bf R}$.
Furthermore, it is evident that $C$, being proportional to the unit element, commutes with all elements of $G$, thus it lies in the center ${\cal Z}(G)$.
Finally, because $N$ clears the denominator of $L({\bf R})$ for all representations, we have $C^N = 1$.
Since by assumption, $N$ is chosen to be the smallest integer such that $N \times L \in \bbZ$, it means that $C$ generates an order $N$ subgroup, i.e., a $\bbZ_N \subset {\cal Z}(G)$.

\subsubsection{Action of the center on F-theory representations}

So far, we have used the explicit form \eqref{eq:fractional_coefficients_shioda} of the coefficients $\lambda_k$ to construct the a $\bbZ_N$ subgroup of the center ${\cal Z}(G)$ associated with a Mordell--Weil generator (free or torsional) $\sigma$.
However, the coefficients $\lambda_k$ also satisfy the integrality condition \eqref{eq:integrality_condition_shioda_map} \cite{Mayrhofer:2014opa}.

For a torsional section $\sigma$, the resulting constraint \eqref{eq:integrality_condition_torsion} implies immediately the integrality of $L({\bf R})$ \eqref{eq:capital_L_definition}.
As a result, we see that the action \eqref{eq:central_element_non-ab_definition} of the center generated by $C$ must be trivial on any representation ${\bf R}$ that is realized in the F-theory geometry!
This means that the gauge group is not $G$, but $G / \langle C \rangle \cong G / \bbZ_N$.

In case the section $\sigma$ is a free Mordell--Weil generator, we have to slightly modify the central element $C$ \eqref{eq:central_element_non-ab_definition}.
First, because $\sigma$ gives rise to a $\fku(1)$, we need to consider representations of the group $U(1) \times G$.
These are specified, in addition to the non-abelian representation ${\bf R}_\fkg$, by the charge $q$.
However, because $U(1)$ only has one-dimensional (irreducible) representations, the representation space of $(q, {\bf R}_\fkg)$---being the tensor product of the two representations $q$ and ${\bf R}_\fkg$---is isomorphic to the representation space of ${\bf R}_\fkg$.
The action of an element $(\exp(2\pi \, i \, \alpha), g) \in U(1) \times G$ is then given by
\begin{align}
	(q , {\bf R}_\fkg) \cong {\bf R}_\fkg \ni {\bf w} \mapsto \left[ e^{2\pi\, i \, q\, \alpha} \otimes \rho(g) \right] {\bf w} = e^{2\pi\, i \, q\, \alpha} \times (\rho(g) \, {\bf w}) \, ,
\end{align}
where $\rho(g)$ is the ${\bf R}_\fkg$-representation of $g$.

With this short interlude, we now define a central element $\tilde{C}$ of $U(1) \times G$ via its action on representation spaces $(q, {\bf R}_\fkg)$:
\begin{align}
	\begin{split}
		{\bf w} \mapsto \tilde{C} \, {\bf w}  : = & \left[ e^{2\pi \, i \,q} \otimes \exp(-2\pi\,i\,L({\bf R}_\fkg) ) \times \mathbb{1}_{\bf R_\fkg} \right] \, {\bf w} \\
		 = & \, \exp\left( 2\pi\,i \left[ q - L({\bf R}_\fkg) \right] \right) \, {\bf w} \, .
	\end{split}
\end{align}
Again, $\tilde{C}$ is obviously in the center, because it commutes with any element of $U(1) \times G$.
Furthermore, we recall that the $U(1)$ charge $q$ is also at most $N$-fractional, because the only non-integer contributions it can receive come again from the coefficients $\lambda_k$ in the Shioda map \eqref{eq:Shioda-map}.
This means that $\tilde{C}^N = 1$, and hence $\langle C \rangle \cong \bbZ_N \subset U(1) \times G$.
Finally, we see the result of the integrality condition \eqref{eq:integrality_condition_free}, which implies $q - L({\bf R}_\fkg) \in \bbZ$.
Therefore, similar to the torsional case, we arrive at the conclusion that $\tilde{C}$ must act trivially on all representations that are realized geometrically in F-theory.
In other words, the global structure of the gauge group is
\begin{align}
	\frac{U(1) \times G}{\langle \tilde{C} \rangle} \cong \frac{U(1) \times G}{\bbZ_N} \, .
\end{align}

It should be noted that recently a ``magnetically'' dual derivation of the gauge group structure has been presented by identifying the so-called cocharacter lattice with a sublattice of the fourth homology group.
An explanation of this intricate result is beyond the scope of these lectures, and we refer the interested reader to the original publication \cite{Monnier:2017oqd}.

\subsubsection{Example: Standard Model gauge group in F-theory}\label{sec:example_SM_XF11_global_group}

The above rather formal discussion has direct relevance for F-theory model building, because it is believed that the Standard Model gauge group has a non-trivial global gauge group structure:
\begin{align}\label{eq:SM_global_gauge_group}
	G_\text{SM} = \frac{SU(3) \times SU(2) \times U(1)}{\bbZ_6} \, .
\end{align}
It turns out that this structure is naturally realized in toric F-theory constructions of the Standard Model \cite{Klevers:2014bqa, Cvetic:2015txa}.
The simplest of these constructions is a given by a hypersurface, whose elliptic fiber is embedded into a toric surface, which described by one of the 16 reflexive 2D polygons.
Explicitly, the hypersurface polynomial reads
\begin{align}\label{eq:XF11_hypersurface}
	p = s_1e_1^2e_2^2e_3e_4^4u^3 + s_2e_1e_2^2e_3^2e_4^2u^2v + s_3e_2^2e_3^3uv^2 + s_5e_1^2e_2e_4^3u^2w + s_6e_1e_2e_3e_4uvw + s_9e_1vw^2 \, ,
\end{align}
where the $s_i$ are sections of various line bundles over the base.
The toric divisors, i.e., the vanishing loci of the coordinates $(u,v,w)$ and $e_i$, restrict to various exceptional divisors and rational sections, which give rise to the Standard Model gauge symmetries when compactifying F-theory on $Y = \{p=0\}$.
Specifically, the Cartan divisor of the $\mathfrak{su}(2)$ is (the restriction of) the divisor $E_1^{\mathfrak{su}(2)} := [e_1]$, whereas the $\mathfrak{su}(3)$ Cartans are the divisors $E_1^{\mathfrak{su}(3)} := [e_2]$ and $E_2^{\mathfrak{su}(3)} := [u]$.
Meanwhile, it is easy to check that the toric divisors $[v]$ and $[e_4]$ restrict to rational sections on the hypersurface,
\begin{align}\label{eq:XF11_sections}
\begin{split}
	\sigma_0 = \{p= 0\} \cap \{v = 0\} : & \quad [u:v:w:e_1:e_2:e_3:e_4] = [1:0:s_1:1:1:-s_5:1] \, ,  \\
	\sigma_1 = \{p=0\} \cap \{e_4=0\} : & \quad [u:v:w:e_1:e_2:e_3:e_4] = [s_9 : 1: 1: -s_3 : 1:1:0] \, , \\
\end{split}
\end{align}
of which we chose to identify the zero section with $\sigma_0$.
Note that we have used some of the projective scalings to set certain coordinates to 1.

One immediately sees that the zero section $\sigma_0$ does not intersect either of the Cartan divisors, since their coordinates are set to 1 in \eqref{eq:XF11_sections}.
On the other hand, the section $\sigma_1$ intersects the $\bbP^1$-fibers of the $\mathfrak{su}(3)$ divisor $[u]$ and the $\mathfrak{su}(2)$ divisor $[e_1]$.
This means that the coefficients $\lambda_k$ \eqref{eq:fractional_coefficients_shioda} in the Shioda map of $\sigma_1$ give rise to the following divisor dual to the $\fku(1)$:
\begin{align}
	\varphi(\sigma_1) = [\sigma_1] - [\sigma_0] + \frac{1}{2} [e_1] + \frac13 ([e_2] + 2[u]) + D_B \, ,
\end{align}
where the pull-back part $D_B$ is the projection term in \eqref{eq:Shioda-map} that is irrelevant for our discussion.
Note that the smallest common denominator of the Shioda map is 6, hence the corresponding central element is of order 6.
In fact, this is the full center of the non-abelian part of the gauge group: ${\cal Z}(SU(3) \times SU(2)) = {\cal Z}(SU(3)) \times {\cal Z}(SU(2)) = \bbZ_3 \times \bbZ_2 = \bbZ_6$.
Following the above discussions, this discrete group is identified with a subgroup of the $U(1)$, such that the global gauge group of the F-theory compactification on the hypersurface \eqref{eq:XF11_hypersurface} is precisely the Standard Model gauge group
\begin{align}
	\frac{SU(3) \times SU(2) \times U(1)}{\bbZ_6} \, .
\end{align}

\subsubsection{The global gauge group as charge constraint and swampland criterion}

In the derivation of the global gauge group of F-theory, the key feature is the integrality condition \eqref{eq:integrality_condition_free}, which on its own is a condition on the $\fku(1)$ charges of non-abelian representations.
In fact, the global gauge group structure is nothing else than such a set of conditions.
For example, the Standard Model gauge group structure \eqref{eq:SM_global_gauge_group} simply means that states in the $({\bf 3,1})$ representation have $U(1)$ charge $\frac13$ mod $\bbZ$, while $({\bf 1,2})$ states have charge $\frac12$ mod $\bbZ$.
Meanwhile, bifundamentals have charge $\frac16$ mod $\bbZ$, and $SU(3) \times SU(2)$ singlets have integral charges.


Field theoretically, statements about $\fku(1)$ charges like these are of course only sensible if one specifies the normalization of the $\fku(1)$.
From that point of view, the only relevant fact is that $\fku(1)$ charges are quantized, and the exact unit of charge quanta is unphysical.
However, in F-theory there is a natural charge quantization, which is inherited from the lattice structure of the Mordell--Weil group, see the discussion of section \ref{sec:lattice_embedding}.
Because in F-theory, matter states arise from holomorphic curves whose intersection numbers with the Shioda divisor gives the charge, the charge quantization of F-theory is naturally given by the fact that also holomorphic curves form a lattice.\footnote{
Concretely, the lattice is the second homology $H_2(Y, \bbZ)$ with integer coefficients.
A representative in there is an integer linear combination of irreducible curves which can be wrapped by M2-branes (possibly multiple times).
The coefficients have to be integral because an M2-brane cannot wrap a fraction of an irreducible curve.
}
Note that in the normalization $\varphi(\sigma) = 1\times [\sigma] + ...$ of the Shioda map \eqref{eq:Shioda-map}, the charge quantization is not necessarily in terms of integers.
In fact, we have argued above that the fractional charges of matter in non-trivial non-abelian representations have important physical consequences.
However, the analysis also shows that in this normalization, the $\fku(1)_\sigma$ charges associated with a free Mordell--Weil generator $\sigma$ of any matter representation $(q_\sigma, {\bf R})$ under $\fku(1) \times \fkg$ satisfy \eqref{eq:integrality_condition_free}:\footnote{
The connection between the coefficients $\lambda_i$ of the Shioda map and the distribution of $\fku(1)$ charges has been noticed and classified for specific examples in \cite{Braun:2013nqa, Kuntzler:2014ila, Lawrie:2015hia, Grimm:2015wda}, although without relating it to the global gauge group structure or exploring its consequences as a possible swampland criterion.
}
\begin{align}\label{eq:charge_constraint_compact}
	q({\bf R}) = L({\bf R}) \mod \bbZ \, .
\end{align}
From this, one immediately arrives at the conclusion that for two matte representations $(q_\sigma, {\bf R}_1)$, $(\tilde{q}_\sigma, {\bf R}_2)$ one has
\begin{align}\label{eq:charge_constraint_difference}
	{\bf R}_1 = {\bf R}_2 \quad \Longrightarrow \quad q_\sigma - \tilde{q}_\sigma \in \bbZ \, .
\end{align}

We claim that this statement is non-trivial in the sense that not all consistent quantum field theories satisfy it.
Phrased differently, it is a criterion that can be used to distinguish low energy limits of string theory from the ``swampland'' \cite{Vafa:2005ui, Ooguri:2006in}, i.e., consistent QFTs without a consistent UV completion including gravity.
However, the statement can only be made with a reference to a chosen normalization of the $\fku(1)$, which for our argument is determined by the Shioda map \eqref{eq:Shioda-map}.
Since the normalization is unphysical, a valid question is if there is any way to test this condition from a purely field theoretic perspective.
After all, as long as the charges are quantized, one can always rescale the $\fku(1)$ such that the charge differences between any matter representations are integral.
Therefore, we first some kind of ``measure stick'' to establish the geometrically preferred normalization in terms of the Shioda map from just field theory data.

We proposed in \cite{Cvetic:2017epq} that non-abelian singlet states provide such measure sticks.
The reason is that, first of all, the charges of such states in F-theory are always integral in the geometrically preferred normalization, cf.~\eqref{eq:charge_constraint_compact}.
Furthermore, it was conjectured in \cite{Morrison:2012ei} and subsequently observed in all explicitly constructed models, that charges of massless singlets---again measured in with the Shioda map \eqref{eq:Shioda-map}---span the full integer lattice $\bbZ^r$, where $r$ is the rank of the Mordell--Weil group, i.e., the number of independent $\fku(1)$s.
Any change of the normalization (i.e., a non-unimodular transformation on the $r$ $\fku(1)$s) would not preserve this property. 
Therefore, one can determine from a purely field theoretic perspective the correct charge normalization by inspecting the charge lattice of spanned by the singlets.

Assuming the validity of the conjecture, we can demonstrate that the condition \eqref{eq:charge_constraint_difference} is stronger than the pure field theory consistency conditions of anomaly cancellation, which are particularly strong for 6D supergravity theories \cite{Kumar:2010ru,Park:2011wv}.
However, we can come up with an anomaly free 6D theory with no tensor multiplets, which nevertheless violates the charge condition \eqref{eq:charge_constraint_difference}:
\begin{align}
\begin{split}
	\text{gauge algebra}& : \, \mathfrak{su}(2) \times \fku(1) \, ,\\
	\text{massless spectrum}&: \, (10 \times {\bf 3}_0) \, \oplus \, (64 \times {\bf 2}_\frac12 ) \, \oplus \, (8 \times {\bf 2}_1 ) \, \oplus \, (24\times {\bf 1}_1) \, \oplus \, (79 \times {\bf 1}_0) \, .
\end{split}
\end{align}
If one rescaled the $\fku(1)$ normalization by 2, then the charges of $\mathfrak{su}(2)$ doublets would satisfy \eqref{eq:charge_constraint_difference}, but this would violate the conjecture that the charges of singlets span $\bbZ$.

\subsection{Gauge enhancement and higher index representations}

From the physics perspective, one can imagine unHiggsing, i.e., enhancing one or several $\fku(1)$ symmetries into a non-abelian gauge algebra.
The geometric description of that phenomenon corresponds to placing the rational sections on special positions on the generic fiber \cite{Morrison:2012ei, Morrison:2014era, Mayrhofer:2014opa, Cvetic:2015ioa, Klevers:2016jsz, Baume:2017hxm}.

We have already seen one such example in section \ref{sec:example_restricted_Tate} in form of the $U(1)$-restricted Tate model:
 \begin{align}\label{eq:U1-restricted_weierstrass_2}
	\begin{split}
		f & = \frac{a_1\,a_3}{2} + a_4 - \frac{1}{48} \, (a_1^2 + 4\,a_2)^2 \, , \\
		g & = \frac{1}{864} \left( (a_1^2+4\,a_2)^3 + 216\,a_3^2 - 36\,(a_1^2 + 4\,a_2)\,(a_1\,a_3 + 2\,a_4) \right) \, .
	\end{split}
\end{align}
The elliptic fibration has a section with coordinates \eqref{eq:weierstrass_coord_U1-restricted} generating a rank 1 Mordell--Weil group.
This changes when we set $a_3 =0$ globally which, as argued in section \ref{sec:example_restricted_Tate}, turns the section to be 2-torsional.
As a consequence, the discriminant of the Weierstrass model \eqref{eq:U1-restricted_weierstrass_2} factorizes:
\begin{align}
	\Delta = 4\,f^3 + 27\,g^2 \quad \stackrel{a_3 =0}{\longrightarrow} \quad \frac{1}{16}  a_4^2 \, \left(4\,a_4 - \left(a_2 + \frac{a_1^2}{4} \right) \right) \, .
\end{align}
This indicates the presence of an $\mathfrak{su}(2)$ gauge algebra over $\{a_4=0\}$, to which the $\fku(1)$ has been enhanced.
Physically, one can also understand this as reversing a Higgs mechanism, in which the non-abelian algebra is broken to its Cartan subalgebra by giving vev to a hypermultiplet in the adjoint representation.
Furthermore, we also know from the previous discussion that the Mordell--Weil group being $\bbZ_2$ implies that the full non-abelian gauge group is $SO(3) = SU(2)/ \bbZ_2$.
This is also reflected by looking at the codimension singular fibers of the tuned geometry, which does not give rise to any fundamental representations of the $\mathfrak{su}(2)$.
This example of unHiggsing has already been studied in \cite{Mayrhofer:2014opa}.
More intricate examples of gauge enhancement by tuning sections to become torsional have been analyzed in \cite{Baume:2017hxm}.

Another way of geometrically altering the Mordell--Weil group is to collide two independent sections, i.e., to tune them such that they sit on top of each other.
For the restricted Tate model \eqref{eq:U1-restricted_weierstrass_2}, such a deformation is not possible, because the only independent sections are the zero section at $[x:y:z] = [1:1:0]$ and the generating section \eqref{eq:weierstrass_coord_U1-restricted}, and the $z$-coordinate of the latter cannot be tuned to zero.
However, the so-called Morrison--Park model \cite{Morrison:2012ei}, which in some sense is the prototype of F-theory models with $\fku(1)$s, can be geometrically unhiggsed this way.
The Weierstrass functions of this model are given by
\begin{align}\label{eq:Morrison-Park_Weierstrass}
	\begin{split}
		f & = c_1\,c_3 - \frac13\, c_2^2 - b^2\,c_0 \, ,\\
		g & =  -c_0\,c_3^2 +\frac13 \, c_1\,c_2\,c_3 - \frac{2}{27} \, c_2^3 + \frac23 \, b^2\,c_0\,c_2 - \frac14 \, b^2\,c_1^2 \, ,
	\end{split}
\end{align}
with the generating rational section at
\begin{align}
	[x:y:z] = \left[ c_3^2 - \frac23 \, b^2\,c_2  :  -c_3^3 + b^2\,c_2\,c_3 - \frac12 \, b^4\,c_1  :  b \right] \, .
\end{align}
One sees immediately that tuning the coefficient $b$ to 0 identifies this section with the zero section.
Physically, this enhances the $\fku(1)$ again to an $\mathfrak{su}(2)$ algebra.
Unlike the previous unHiggsing example via Mordell--Weil torsion, this enhanced model has gauge group $SU(2)$.
Consistently, the spectrum now also contains doublet states.

In this example, the $\fku(1)$ model to begin with had singlets with charge 1 and 2.
By enhancing the abelian symmetry into an $\mathfrak{su}(2)$, the charge 1 and 2 states become ${\bf 2}$ resp.~$\bf 3$ representations. i.e., the $\fku(1)$ charges are mapped directly onto the Cartan charges of the non-abelian representations.
Repeating the same tuning process for a $\fku(1)$ model with charge 3 singlets, it was able to construct an F-theory model with the three-index symmetric representation, i.e., the ${\bf 4}$ of $\mathfrak{su}(2)$ \cite{Klevers:2016jsz, Klevers:2017aku}.
Going beyond rank 1, one can also enhance a model with $\fku(1)^2$ model and charge $(2,2)$ singlets into an $SU(3)$ theory with the two-index symmetric representation ${\bf 6}$ by colliding all three independent sections \cite{Cvetic:2015ioa}.
However, these two examples are so far the only two explicit F-theory realizations of higher index symmetric matter representations.
Recently, a $\fku(1)$ model with charge 4 singlets has been constructed \cite{Raghuram:2017qut}, but a similar attempt of gauge enhancement led to a larger gauge group with higher charge adjoints instead of the ${\bf 5}$ representation of $\mathfrak{su}(2)$.
It seems there is some arguments in terms of the fiber structure of F-theory that forbids this and other higher index representations in F-theory, at least in terms of Kodaira fibers \cite{Klevers:2017aku}.
On the other hand, given that F-theory is dual to heterotic string theory, where it has been known for a long time how to engineer higher representations on orbifolds, there must be some dual description also in terms of elliptic fibrations.
Recently, it has been argued that such constructions, at least in 6D, are likely to always involve non-minimal singularities \cite{Ludeling:2014oba, Cvetic:2018xaq}, and hence not excluded by the other arguments.
However, it remains an open question how to systematically construct F-theory models where one can explicitly show the presence of higher index representations.

\section{Discrete Abelian Symmetries in F-theory}\label{sec:discrete}

At the end of the previous section, we have discussed gauge enhancing $\fku(1)$ symmetries into non-abelian ones, and presented the geometric description in terms of colliding multiple rational sections.
The resulting elliptic fibration has a smaller Mordell--Weil group (concretely, the rank is lower), but has additional exceptional divisors in codimension one.
The other direction, namely Higgsing the $\fku(1)$ to a discrete subgroup, is an equally interesting question.
It turns out that to fully understand the process in F-theory, one has to go beyond elliptically fibered geometries and allow fibrations without rational sections.

\subsection{Discrete symmetries in field theory}

In order to know what physical features of discrete symmetries we need to find in a geometric description, we shall first briefly review the field theoretic description of discrete abelian symmetries and their origin in terms of a broken $\fku(1)$.
More details of this discussion can found in any standard textbook (e.g., \cite{peskin1995introduction}), and here we will focus only on the relevant parts.

Let us begin with a comples scalar field $\phi$ with charge $n \in \mathbb{N}$ under a $\fku(1)$ gauge field $A$.
The kinetic term of this scalar field in the Lagrangian is
\begin{align}\label{eq:lagrangian_u1}
	{\cal L} \supset D_\mu \phi \, \overline{D^\mu \phi} = (\partial_\mu \phi + i \, n \, A_\mu \, \phi)\, (\partial^\mu \bar{\phi} - i \, n \, A^\mu \, \bar{\phi}) \, .
\end{align}
By giving a vacuum expectation value (vev) $v = \langle \phi\rangle$ to $\phi$, i.e., $\phi = \frac{1}{\sqrt{2}}(v + h) \,e^{i \, c}$, the kinetic term gives rise to the so-called St\"uckelberg Lagrangian:
\begin{align}\label{eq:Stueckelberg_lagrangian}
	(D\phi)^2 \stackrel{\langle \phi \rangle}{\longrightarrow} \frac{v^2}{2} \left( \partial c + n \,A \right)^2 + ... \, .
\end{align}
The real part $h$ of the perturbations of $\phi$ around the vev would corresponds to the Higgs boson, which is not part of the massless spectrum and will be hence ignored in the subsequent discussion.
The scalar $c$ on the other hand is the massless Goldstone boson, and furthermore enjoys a shift symmetry $c \cong c + 2\pi$, simply because it is a phase which is only defined up to a periodic identification.
Scalar fields with shift symmetry are usually called axions.
In the case of the St\"uckelberg axion, its shift symmetry is gauged by the $\fku(1)$ symmetry.
Namely, the Lagrangian \eqref{eq:Stueckelberg_lagrangian} is invariant under
\begin{align}\label{eq:gauged_shift_symmetry}
	A \rightarrow A + \partial \alpha \, , \quad c \rightarrow c - n \, \alpha \, .
\end{align}

In representation theory, $c$ furnishes a so-called affine, or non-linear representation of $\fku(1)$.
Abstractly, whenever there are degrees of freedom transforming non-linearly under a symmetry transformation, this symmetry is said to be spontaneously broken.
In the case of the $\fku(1)$ gauge symmetry, a more physical way to see the breaking is to exploit the transformation \eqref{eq:gauged_shift_symmetry} to completely gauge away the axion in \eqref{eq:Stueckelberg_lagrangian} ($\alpha = c/n$), yielding a mass term for the vector field with mass $m^2 = n^2 v^2/2$.
In the context of the Higgs mechanism, this effect is often referred to as the Goldstones being ``eaten'' by the massive gauge bosons.

While the mass term makes the spontaneous symmetry breaking mechanism physically very intuitive, the abstract classification via linearly vs.~non-linearly realized transformations explains very easily why there is still a discrete part of the $\fku(1)$ symmetry left intact.
Namely, whenever $\mathbb{N} \ni n > 1$, a subset of transformations \eqref{eq:gauged_shift_symmetry} with $\alpha = 2\pi \frac{k}{n}$, where $k \in \bbZ$, act trivially (and, hence, linearly) on $c$ because of the shift symmetry $c \cong c + 2\pi$!
The corresponding subgroup of the $U(1)$ is $exp(i \, \alpha)$, i.e., $\bbZ_n$.
Other matter fields that were originally charged under the $\fku(1)$ now transform non-trivially under this discrete subgroup.
One can assign a representation to them, which is just the $\fku(1)$ charge mod $n$.
An important physical implication of such discrete symmetries is that they can forbid Yukawa couplings in 4D.
Thus, they provide a very attractive way to construct selection rules without having to introduce exotic gauge bosons, since the gauge bosons responsible for this symmetry are rendered massive.

\subsection{Geometric description of discrete symmetries in M-theory}

To describe non-abelian symmetries geometrically, we have to remind ourselves that the geometric phase of F-theory is described via duality to M-theory.
Thus, it seems to be natural to first understand discrete symmetries in M-theory.

There, it is known \cite{Grimm:2010ez, Grimm:2011tb} that massive $\fku(1)$ gauge fields $A$, correspond to expansions of the M-theory three-form $C_3$ along non-harmonic two-forms of the compactification space $Y$.
Concretely, we have:
\begin{align}\label{eq:non_harmonic_form}
	\text{d} \omega_2 = n \, \eta_3 \, .
\end{align}
To consistently incorporate this relation in the low energy physics, we must include $\eta_3$ in the Kaluza--Klein expansion,
\begin{align}
	C_3 = A \wedge \omega_2 + c\,\eta_3 + ... \, .
\end{align}
Then, the dimensional reduction of the kinetic term $\text{d}C_3 \wedge \ast \text{d} C_3$ in 11D precisely yields the St\"uckelberg mass term \eqref{eq:Stueckelberg_lagrangian}.
For $n \neq 0,1$, the non-harmonic forms \eqref{eq:non_harmonic_form} give rise to a non-trivial torsion class in integer cohomology
\begin{align}
	\eta_3 \in \text{Tors}\left( H^3 (Y, \bbZ) \right) \, .
\end{align}
The corresponding discrete symmetry uplifts directly to F-theory via the M-/F-theory duality \cite{Braun:2014oya, Morrison:2014era, Mayrhofer:2014laa}.

One practical problem with torsional cohomology is that it is notoriously hard to detect in a given geometry.
However, there is another geometric consequence of discrete symmetries which is more tractable.
This arises from having massless matter states which are only charged under the discrete symmetry, i.e., the massive $\fku(1)$ field $A$.
These arise in M-theory from M2-branes wrapping collapsed 2-cycles $\Gamma$ inside the Calabi--Yau $Y$, which cannot be blown-up while keeping the manifold $Y$ a K\"ahler space \cite{Braun:2014oya, Braun:2014nva}.
Field theoretically, this means that one cannot give a mass to these states in M-theory on $Y$ by going onto the Coulomb branch without breaking supersymmetry.
Hence, if we restrict ourselves to supersymmetric compactifications, i.e., internal spaces $Y$ which are Calabi--Yau manifolds, we necessarily have to have ``terminal singularities'' (such that cannot be blown-up in a K\"ahler manifold) on $Y$.
Terminal singularities can oftentimes be detected straightforwardly on a given manifold, and have been recently studied carefully on Calabi--Yau threefolds, together with their enumeration in terms of 6D anomalies in F-theory \cite{Arras:2016evy}.
There is however still a drawback of using terminal singularities to detect $\bbZ_n$ symmetries, since they only signal the presence of matter charged under a massive vector field, but neither its charge nor the remnant discrete symmetry of the field theory can be determined.

It turns out that the most convenient description of an F-theory model with discrete symmetry is a manifold $\bbY$ that neither has torsion nor terminal singularities.
In fact, it is not even elliptically fibered.
Rather, the manifold $\bbY$ is a \textit{genus-one} fibration with so-called multi-sections.
We will explain in the following how these spaces differ from elliptic fibrations, and how these differences can circumvent terminal singularities, allowing an easy way to determine the matter charges under the discrete abelian symmetry.
The crucial insight here will be that discrete symmetries in F-theory does not necessary imply discrete symmetries in the dual M-theory.

\subsection{F-theory on genus-one fibrations}

\begin{figure}[!ht]
\centering
	\def\svgwidth{.8\hsize} 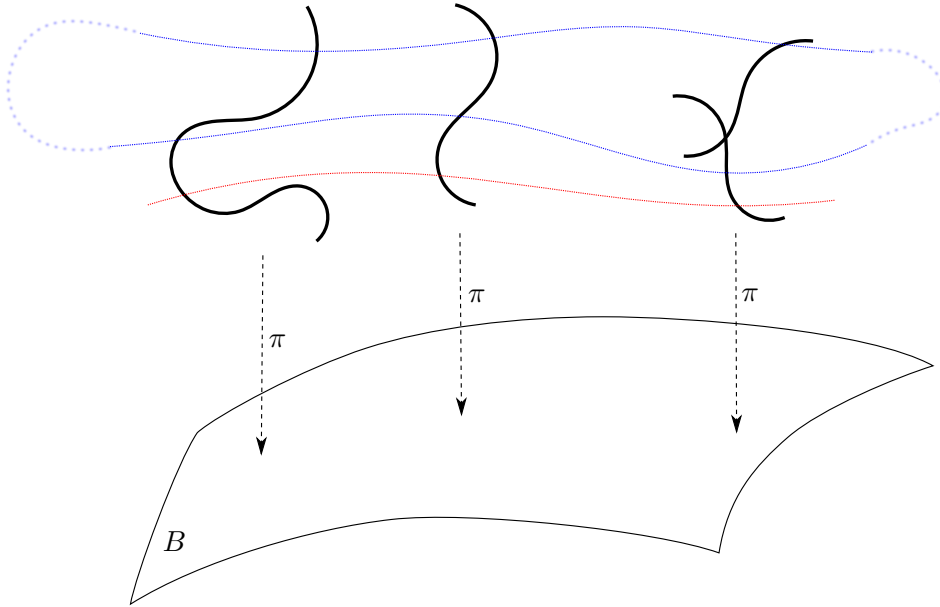 
	\caption{A rational or 1-section (red) intersects each fiber of a genus-one fibration $\pi: Y \rightarrow B$ exactly once.
	A bi- or 2-section (blue) intersects each fiber in two points.
	Globally there is a monodromy exchanging these two points.}\label{fig:section_vs_multi_section}
\end{figure}

As has been extensively discussed in recent works \cite{Braun:2014oya, Morrison:2014era, Anderson:2014yva, Mayrhofer:2014haa, Klevers:2014bqa, Garcia-Etxebarria:2014qua,  Mayrhofer:2014laa, Cvetic:2015moa, Lin:2015qsa, Kimura:2016gxw, Kimura:2016crs}, F-theory can be defined on a Calabi--Yau space $\bbY$ that is torus fibered over a K\"ahler base $B$, but has no rational section, that is, it is not elliptically fibered.
We will follow the nomenclature that has been established in the literature and call these genus-one fibrations.
In genus-one fibrations, there always exists a minimal $n \in \mathbb{N}$ such that there is a divisor $s^{(n)}$ of $\bbY$ which is an $n$-fold cover over $B$.
Because such a divisor intersects the generic torus fiber $n$ times, it is oftentimes called a $n$- or multi-section.
In this setting, a rational section would be a 1-section.
The difference between the two is that a section marks a single point on the generic fiber, hence can be thought as a map from the base into the total space of the fibration.
An $n$-section on the other hand associates a collection $\{p_l\}_{l\leq n}$ of $n$ points on the fiber over a single point.
If one singles out one of these points $p_1$ and traces its movement along the fibers as one continuously moves the point in the base, then one observes that for certain closed paths in the base, i.e., where one ends up in the same fiber, the marked point becomes one of the other $n$ points, say $p_2$.
For a rational section in an elliptic fibration, this can never happen.
For genus-one fibrations however, only a collection of $n$ points can be invariant under such monodromy actions.
In figure \ref{fig:section_vs_multi_section}, we have illustrated a bisection and put it in contrast to an ordinary 1-section.
As we will explain now, these geometries provide a different, but physically equivalent description of discrete abelian symmetries in F-theory.

Like the case of elliptic fibrations, the geometry itself only has a direct interpretation in M-theory.
For our purposes, we need to identify the M-theory compactification as a circle reduction of a theory, which by definition is \textit{the} F-theory on $\bbY$.
As a circle reduction, M-theory on $\bbY$ necessarily needs to have a massless $\fku(1)$ which accounts for the Kaluza--Klein $\fku(1)$.
In this case, it is provided by the divisor class of the multi-section $s^{(n)}$.
However, as already mentioned before, the genus-one fibration $\bbY$ is in general smooth and has no torsional cohomology.
This begs the question how the discrete symmetries, which are clearly absent in the M-theory compactification on $\bbY$, manifest themselves in F-theory on $\bbY$.

The subtleties lie in the process of circle compactification.
Concretely, when compactifying a field theory with a vector field $A$ in $F = M+1$ dimensions on an $S^1$, one can turn on a flux along the circle,
\begin{align}
	\xi = \int_{S^1} A \, .
\end{align}
If the vector field is associated with an unbroken gauge symmetry in $F$ dimensions, $\xi$ is referred to as a holonomy, and parametrizes a gauge transformation when encircling the $S^1$ once.
For a continuous symmetry with algebra $\fkg$, $\xi$ is a continuous parameter taking value in the Cartan subalgebra of $\fkg$, which, if non-zero, breaks the gauge symmetry to a rank($\fkg) = r$ subalgebra of $\fkg$.
Generically, this is simply the Cartan subalgebra $\fku(1)^r$.
Geometrically, changing the values of $\xi$ continuously changes the sizes of the fibral $\bbP^1$ components of codimension one and two reducible fibers.\footnote{
More precisely, the parameter $\xi$ are coordinates on the Coulomb branch of the theory in $M$ dimensions.
Geometrically, it corresponds to the extended K\"ahler cone of the Calabi--Yau.
}
In that sense, different $\xi$'s define different manifolds, which however are connected by continuous deformations.

However, for a discrete $\bbZ_n$ symmetry with a massive gauge field $A$ in $F$ dimensions, the allowed holonomies are discrete.
Hence, we do not expect that the compactified theories in $M$ dimensions with different values of $\xi$ are connected continuously.
Indeed, the picture that has emerged over the last few years is that both multi-section geometries $\bbY$ and elliptic fibrations  $Y$ with torsional cohomology and terminal singularities can describe the same F-theory in $F$ dimensions.
Their apparent difference is reflecting different choices of the discrete holonomy $\xi$ when we compactify on a circle to go down to M-theory.

If the holonomy is trivial, then the discrete symmetry descends straightforwardly to $M$ dimensions.
This is the situation when we compactify M-theory on an elliptic fibration with torsional cohomology.
The zero section of the fibration gives rise to the KK-$\fku(1)$, and the torsional cohomology encodes the $\bbZ_n$ symmetry.
On the other hand, if the holonomy is non-trivial, it turns out that the Kaluza--Klein reduction along the fluxes $S^1$ gives rise to a kinetic mixing term between the KK-$\fku(1)$ and the massive $\fku(1)_m$ in the Lagrangian for the $M$-dimensional theory \cite{Anderson:2014yva, Mayrhofer:2014haa, Cvetic:2015moa}.
As a result, the true massless $\fku(1)$ in $M$-dimensions is a linear combination of the KK-$\fku(1)$ and the massive vector field.
This massless linear combination is the $\fku(1)$ which is dual to the divisor class of the multi-section, when we compactify M-theory on a genus-one fibration $\bbY$.

In fact, there is some deep mathematics associated with this physics description.
As noted in \cite{Braun:2014oya}, we can associate to any genus-one fibration $\bbY$ an elliptic fibration $\bbY$ with the same base, which has the same discriminant locus, i.e., they encode the same 7-brane configuration in the base $B$.
$Y$ is the so-called Jacobian fibration associated with $\bbY$, sometimes denoted as $Y = J(\bbY)$.
It turns out that the existence of genus-one fibrations of the same dimension and with the same discriminant locus as an elliptic fibration $Y$ is closely related to the torsional cohomology of $Y$, i.e., discrete symmetries in M- and F-theory.
It has been proven for threefolds $Y$ with no reducible fibers in codimension one that the torsional cohomology is encoded in the so-called Tate--Shafarevich group \cite{MR1242006},
\begin{align}
	\text{Tors} \left( H^3 (Y, \bbZ) \right) \cong \Sh(Y) \, .
\end{align}
This group, whose precise definition is beyond the scope of these notes, appears in the arithmetic geometry of elliptic and genus-one fibrations.
The key property of $\Sh(Y)$ is however, that its element are genus-one fibrations $\bbY$ whose Jacobian are $Y$.
In other words, the Tate--Shafarevich group is the collection of different M-theory vacua, whose F-theory uplift are equivalent, namely a field theory with $ \Sh(Y) = \bbZ_n$ gauge symmetry.
Moreover, the order $n$ of the discrete symmetry is the minimal integer for which there exists a multi-section of that degree.

In practice, explicit construction of the Tate--Shafarevich group in the F-theory literature have only gone as high as $n=3$.
For $n=2$, it is obvious that the geometries $Y$ and $\bbY$, where $\bbY$ has a 2- or bisection, are different elements of $\Sh(Y)$.
However, the important observation of \cite{Cvetic:2015moa} is that elements of $\Sh(Y)$ are in general specified by more than just a geometry.
This becomes crucial in the case of $n=3$, where the two non-trivial elements of $\Sh(Y) = \bbZ_3$ both share the same underlying geometry $\bbY$, which has a tri-section, but differ by these additional, more subtle data.\footnote{These data, among others, include the specification of a map $f: \bbY \rightarrow Y = J(\bbY)$, which can be defined in two different ways in case $\Sh(Y) = \bbZ_3$.}
An interesting question would be to analyze if for higher $n$, also the underlying geometry can differ between different non-trivial elements of $\Sh(Y)$.
The natural candidate would be a model with $n=4$, for which there exists an explicit construction of a genus-one fibration with a four-section \cite{Braun:2014qka, Oehlmann:2016wsb}.
It is unclear at this point however, if there might be one or more geometrically non-isomorphic genus-one fibration that form the full $\Sh(Y) = \bbZ_4$ group.

\subsubsection{Discretely charged matter in genus-one fibrations}

We have claimed earlier that it is easier to read off the discrete gauge symmetry as well as matter charges under it in a genus-one fibration $\bbY$, than in its Jacobian $Y$.
The order $n$ of the discrete symmetry, as already seen earlier, corresponds to the minimal degrees of multi-sections in $\bbY$.
Concerning the matter states, we analyze the fiber of $\bbY$ over the codimension two loci of $B$, where the Jacobian fibration $Y$ had terminal singularities.
The justification is that, since both encode the same 7-brane configuration, the charged matter have to be localized at the same points of the type IIB compactification space $B$.

It turns out that in $\bbY$, there are no terminal singularities in these fibers.
Instead, the fibers are of Kodaira-type I$_2$, meaning it consists of two $\bbP^1$s intersecting each other transversely in two points.
The important observation is that the $n$-section will now intersect each component non-trivially: 
\begin{align}\label{eq:curves_with_discrete_states}
	\begin{split}
		& s^{(n)} \cdot \bbP^1_a = k \, ,\\
		& s^{(n)} \cdot \bbP^1_b = n-k \, .
	\end{split}
\end{align}
To interpret this physically, recall that the $\fku(1)$ dual to the $n$-section is a linear combination of the KK and the massive $\fku(1)_m$. Concretely, it is \cite{Anderson:2014yva, Mayrhofer:2014haa, Cvetic:2015moa}
\begin{align}
	\fku(1)^{(n)} = n\,\fku(1)_\text{KK} - \fku(1)_\text{m} \, .
\end{align}
For states uncharged under the discrete symmetry, i.e., under the massive $\fku(1)_\text{m}$, the intersection number with the $n$-section is just $n$-times the KK-charge.
For $\fku(1)_\text{m}$-charged states, the KK-charge is shifted, and now deviates from being a multiple of $n$.
These are now precisely the case for the states on the $\bbP^1$s satisfying \eqref{eq:curves_with_discrete_states}.
Moreover, note that by just measuring the $\fku(1)^{(n)}$ charge, we can only determine the $\fku(1)_\text{m}$ charge up to multiples of $n$.
This is of course consistent with the fact that the actual gauge symmetry is $\bbZ_n$, i.e., the charges are only defined mod $n$.
The upshot is that we now have an easy way of determining the $\bbZ_n$ charge: it is simply the intersection number of the fibral curve with the $n$-section, taken modulo $n$.
In \eqref{eq:curves_with_discrete_states}, the states would thus have chargers $k$ and $-k$, which consistently form a charge conjugate pair.

When we include non-abelian gauge algebras via reducible fibers in codimension one, then one obtains additional, independent divisors corresponding to the Cartan divisors $E_i$.
Because a multi-section has several ``prongs'' that can intersect several $\bbP^1$ fibers of different exceptional divisors $E_i$, the non-abelian W-bosons would be charged under the discrete symmetry.
To remedy this, one can, similar to the case of massless $\fku(1)$s, define a Shioda-like divisor,
\begin{align}
	\varphi(s^{(n)}) = [s^{(n)}] + \sum_{i,j}   s^{(n)} \cdot \bbP^1_i (C^{-1})_{ij} E_j \, ,
\end{align}
where $C^{-1}$ is again the inverse Cartan matrix of the non-abelian gauge algebra $\fkg$.
Because of the appearance of the inverse Cartan matrix, the charges obtained by computing intersection numbers with $\varphi(s^{(n)})$ are in general fractional.
The interpretation in terms of a discrete charge actually means, that the discrete symmetry is enhanced by an order $m$ subgroup of the center of $\fkg$ to $\bbZ_{n \cdot m}$ \cite{Garcia-Etxebarria:2014qua}.
However, a similar analysis to the case of $\fku(1)$s show that in this situation, there is also a non-trivial charge constraint which induces a non-trivial global gauge group structure of the form \cite{Cvetic:2018ryq}
\begin{align}
	\frac{G \times \bbZ_{n \cdot m}}{\bbZ_m} \, .
\end{align}
Now, if $n$ and $m$ are coprime, then the ``Chinese remainder theorem'' ($\bbZ_{n \cdot m} \cong \bbZ_n \times \bbZ_m$) leads to a cancellation of the enhancing $\bbZ_m$ factor, effectively leading to the ``naive'' global gauge group $G \times \bbZ_n$.
This ``accidental'' cancellation allowed for a somewhat careless treatment of the discrete charges in early phenomenologically motivated F-theory constructions of $SU(5) \times \bbZ_2$ models \cite{Mayrhofer:2014haa, Lin:2015qsa}.
However, later examples with $\mathfrak{su}(2)$ algebras \cite{Klevers:2014bqa, Cvetic:2018ryq} precisely show such an enhancement of the discrete symmetry to a $\bbZ_4$, even though the genus-one fibration had a 2-section.

\subsection{Geometric description of Higgsing}

Even though we have motivated the study of discrete symmetries via the Higgs mechanism at the begin of this section, we have not yet discussed how this process manifest itself in F-theory.
In particular, can we understand the different M-theory vacua, whose geometry differ so significantly, as coming from a single F-theory model with $\fku(1)$, for which there does not seem to be any ambiguities in terms of geometric characterization?
The subtlety is that already in the F-theory model with $\fku(1)$, there were strictly speaking several geometries which differed in M-theory only in their massive spectrum, but not the gauge symmetry.
For simplicity, let us look at an example with $n=2$.
The subtleties that arise for $n=3$ are explained in \cite{Cvetic:2015moa}.

\subsubsection{Higgsing in the Weierstrass model}

The $\fku(1)$ phase of this story is the Morrison--Park model, whose Weierstrass model we have already written down above (see \eqref{eq:Morrison-Park_Weierstrass}).
This theory has a charge 2 singlet, which is geometrically realized as an I$_2$ fiber at $b=c_3 =0$.
Furthermore, there are also charge 1 singlets, again realized as I$_2$ fibers at a different codimension two locus (described by a non-complete intersection $V(I)$ of a complicated ideal $I$).

Now, the Higgs mechanism is described geometrically via a generalized conifold transition \cite{Braun:2011zm, Krause:2012yh, Intriligator:2012ue}.
In order to obtain a $\bbZ_2$ from the $\fku(1)$, we therefore first blow-down the $\bbP^1$ component not intersected by the zero section over the locus $b=c_3=0$, and subsequently deform the geometry to smooth out the singularity.
The smoothing process is described via a complex structure deformation $b^2 \rightarrow 4c_4$ in the Weierstrass equations \eqref{eq:Morrison-Park_Weierstrass}.
As explained in detail in \cite{Mayrhofer:2014laa}, the blow-down process inevitably also shrinks a $\bbP^1$ component over the other locus $V(I)$ which hosts the charge 1 singlets.
But the singularity created in this way is not deformed away through the complex structure deformation.
This way, the resulting geometry $Y$, even though it is still elliptically fibered, now has terminal singularities, sitting precisely at the locus where charge 1 matter states are localized, which now turn into the charged singlets of the $\bbZ_2$.
A more careful analysis \cite{Mayrhofer:2014laa} then also reveals the presence of $\bbZ_2$ torsional cohomology, confirming the discrete symmetry in M-theory on $Y$.

\subsubsection{Higgsing in the toric hypersurface}

As shown in \cite{Morrison:2012ei}, the same $\fku(1)$ theory can be described by a toric hypersurface $X_T$,
\begin{align}\label{eq:MP_weighted_proj}
	w^2 + b \,w\,v^2 = c_0  \, u^4+ c_1 \, u^3 \,v  + c_2 \, u^2 \, v^2 + c_3 \, u \, v^3 \, ,
\end{align}
where the coordinates $[u:v:w]$ are those of a weighted projective space $\bbP_{112}$, and the coefficients $b$ and $c_i$ are functions over the base $B$ of the fibration, which is the same as the base of the Weierstrass model $X_W$, given by \eqref{eq:Morrison-Park_Weierstrass}.
This hypersurface has two rational sections, given by the intersection points with $u=0$:
\begin{align}
	\begin{split}
		&\sigma_0 : [u : v :w] = [0 : 1 : 0] \, , \\
		& \sigma_1 : [u:v:w] = [0 : 1 : -b \, .]
	\end{split}	 
\end{align}
Note that this fibration also has I$_2$ fibers over $b=c_3=0$ and $V(I)$, giving rise to matter charged under the $\fku(1)$ in F-theory.
When we pass over to the Weierstrass model $X_W$, we identify the section $\sigma_0$ with the zero section.
However this map is only a birational equivalence, meaning that $X_W$ and $X_T$ can differ in codimension two and higher.
In this case, the difference is in the K\"ahler and Mori cone structures, i.e., the possibilities how one can shrink and blow-up curves without violating the Calabi--Yau condition of the space.

For the toric hypersurface, the conifold transition that gives a vev to the charge 2 singlets again requires to blow-down a fiber component over $b=c_3=0$ and subsequently deforming away the resulting singularities via $b^2 \rightarrow 4c_4$.
However, because of the different K\"ahler and Mori cone structures, the blow-down now does \textit{not} affect the fibers over $V(I)$ \cite{Mayrhofer:2014laa}.
Consequently, there are no terminal singularities in these fibers after the deformation, which produces a genus-one fibration $\bbY$.

To see the genus-one nature explicitly, we have to make a coordinate redefinition $w = \tilde{w} - \frac12 b \, v^2$, which modifies the left-hand side of \eqref{eq:MP_weighted_proj} to $\tilde{w}^2 - \frac14\,b\,v^4$.
Then, the complex structure deformation $b^2 \rightarrow 4c_4$, with $c_4$ a generic non-square polynomial, yields a new hypersurface,
\begin{align}
	\tilde{w}^2 = c_0\,u^4 + c_1\,u^3\,v + c_2 \,u^2\,v^2 + c_3\,u\,v^3+ c_4\,v^4 \, ,
\end{align}
which does not exhibit any rational section.
However, it does have a bisection, given by the intersection of $u=0$, which marks in any fiber the two points which are roots of the quadratic equation $\tilde{w} = c_4\,v^2$.\footnote{Note that $v=0$ equally defines a bisection that is in the same class as $u=0$. In fact, it is not hard to show using the Riemann--Roch theorem that there are in general $n$ different $n$-sections with the same divisor class.}

Finally, let us remark that the two different geometries $X_W$ and $X_T$ for the $\fku(1)$ theory are connected to each other via a continuous K\"ahler deformation.
However, this connection involves a so-called ``flop'' transition: at some point of the continuous deformation, a curve shrinks to zero size, thus creating a singularity.
This singularity is then resolved by blowing up a different curve.
Physically, the deformation parameter is related to the flux, or holonomy, of the $\fku(1)$ gauge field along the circle in the reduction from F- to M-theory, which before the Higgsing is a continuous parameter.
The two configurations corresponding to either the blown-down phase of $X_W$ or $X_T$ can be thought of as two special values for the $\fku(1)$ holonomy.
Only at these two special values is the complex structure deformation $b^2 \rightarrow 4c_4$ accessible.
However, once we turn on this deformation, then the curves whose volumes changed with the flux parameter are gone from the geometry.
As a consequence, the flux is ``frozen'' to these particular values.
Physically, these two situations are of course precisely the two distinct possibilities of the $\bbZ_2$ holonomy, which are now no longer connected continuously in the M-theory moduli space.
In geometry, we observe these now as the two elements $Y$ and $\bbY$ of the Tate--Shafarevich group $\Sh(Y) = \bbZ_2$.

\section{Application: Global Particle Physics Models}\label{sec:application}

One of the major physical motivation for studying abelian symmetries in F-theory is their importance for particle phenomenology.
While $\fku(1)$ symmetries feature prominently in the Standard Model as the hypercharge, discrete symmetries provide a minimally invasive extension that can serve as a selection rule.
In the following, we will present three examples, each realizing the Standard Model gauge algebra, but with a different extension.
The significance of these models is that they are all globally defined models, i.e., the full compact Calabi--Yau space can be specified.
This is to be distinguished from the early day F-theory model building attempts, which were more restricted to local constructions of GUT models.
One significant advantage over the local treatment is that it is possible to determine consistent $G_4$-flux configurations that generate a chiral spectrum.\footnote{We have collected some basic facts about $G_4$-fluxes in F-theory in appendix \ref{app:fluxes}. For a more comprehensive discussion, see \cite{Weigand:2018rez}.}
Indeed, for all three examples, explicit configurations with low numbers or no chiral exotics have been found.

\subsection{The minimalistic example}

The most natural example is of course to realize just the Standard Model gauge group \cite{Cvetic:2015txa}.
The elliptic fibration for that has already been presented in section \ref{sec:example_SM_XF11_global_group}.
There, we have focused on the rational sections and the codimension one singular fibers, which gave rise to the exact Standard Model gauge group
\begin{align}
	G_\text{SM} = \frac{SU(3) \times SU(2) \times U(1)}{\bbZ_6} \, .
\end{align}
By inspecting the codimension two enhancement, we find that this F-theory model contains the same representations as the Standard Model, which we collect together with their geometric loci in table \ref{tab:poly11_matter}.
\begin{table}[ht]
\begin{center}
\renewcommand{\arraystretch}{1.2}
\begin{tabular}{|c| c| c|}\hline
Representation & Locus & SM-matter \\ \hline
$(\three,\two)_{1/6}$ & $\{ s_3=s_9=0 \}$ & left-handed quarks $Q$ \\ \hline
$(\one,\two)_{-1/2}$ & $ \{ s_3=s_2s_5^2 + s_1(s_1 s_9-s_5s_6)=0 \}$ & lepton $L$ and Higgs $H$ doublets \\ \hline
$(\overline{\three},\one)_{-2/3}$ & $\{ s_5=s_9=0 \}$ & right-handed up-quark $\bar{u}$ \\ \hline
$(\overline{\three}, \one)_{1/3}$ & $\{ s_9= s_3 s_5^2 + s_6(s_1 s_6-s_2 s_5)=0 \}$ & right-handed down-quark $\bar{d}$ \\ \hline
$(\one,\one)_{1}$ & $\{ s_1=s_5=0 \}$ & right-handed electron $e$ \\ \hline
\end{tabular}
\caption{\label{tab:poly11_matter} Charged matter representations under $\mathfrak{su}(3) \times \mathfrak{su}(2) \times \fku(1)$ and corresponding codimension two loci of the minimalistic example.}
\end{center}
\end{table}

To specify a concrete model, one has to specify the base $B$ as well as the divisor classes of the coefficients $s_i$.
As demonstrated in \cite{Cvetic:2015txa}, for the simplest choice of base, namely $B=\bbP^3$, one can find configurations that have consistent $G_4$-flux vacua that leads to the precise chiral Standard Model spectrum, namely three chiral families for each of the matter representations listed in table \ref{tab:poly11_matter}.

A drawback of this model is the lack of selection rules which forbid certain R-parity violating Yukawa couplings, which can generate problematic interactions which are constrained by today's experiments.
For example, because the Higgs and the lepton doublet have the same quantum numbers under the Standard Model, they have to be localized on the same locus in this F-theory model.
As a consequence, it is hard to come up with a mechanism that generates an order one top Yukawa coupling $Q H \bar{u}$, but suppresses the coupling $Q L \bar{u}$ which contributes to proton decay.

To remedy this problem, phenomenologists have come up with various approaches.
One of them is to introduce an additional gauged $\fku(1)$ symmetry, such as $U(1)_{B-L}$ or Peccei--Quinn symmetry.
Therefore, it is also interesting to look at potential F-theory realizations of such extensions to the Standard Model.

\subsection[F-theory models with \texorpdfstring{$\mathfrak{su}(3) \times \mathfrak{su}(2) \times \fku(1)^2$}{SU3 x SU2 x U1 x U1} symmetry]{F-theory models with \boldmath{$\mathfrak{su}(3) \times \mathfrak{su}(2) \times \fku(1)^2$} symmetry}

In order to geometrically engineer a model with two $\fku(1)$s, the elliptic fibration needs to have three independent rational sections.
Such an example is provided by a toric hypersurface where the fiber is embedded inside the surface $\text{Bl}_2 \bbP^2$, that is $\bbP^2$ (with coordinates $[u:v:w]$) blown-up at two points (by $s_0$ and $s_1$) \cite{Borchmann:2013jwa, Cvetic:2013nia, Borchmann:2013hta,Cvetic:2013jta, Lawrie:2014uya}.
The hypersurface polynomial is
\begin{align}\label{eq:hypersurface_two_u1s}
	vw (c_1\,w s_1 + c_2\,v s_0) + u (b_0\,v^2 s_0^2 + b_1\,v w s_0 s_1 + b_2\,w^2 s_1^2) + u^2 (d_0 \,v s_0^2 s_1 + d_1 \, w s_0 s_1^2 + d_2\,u s_0^2 s_1^2) \, ,
\end{align}
where the coefficients $b_i, c_j, d_k$ are again some holomorphic functions over the base.
The three rational sections are given by the intersection of the hypersurface \eqref{eq:hypersurface_two_u1s} with the three toric divisors of the fiber ambient space:
\begin{eqnarray}
	\begin{aligned}
		\sigma_0 & = \{s_0\} & : \quad [u:v:w:s_0:s_1] = [-c_1 : b_2 : 1 : 0 : 1] \, ,\\
		\sigma_1 & = \{s_1\} & : \quad [u:v:w:s_0:s_1] = [-c_2 : 1: b_0 :1 :0] \, , \\
		\sigma_2 & = \{u\} & : \quad [u:v:w:s_0:s_1] = [0 : 1 : 1 : -c_1 : c_2] \, .
	\end{aligned}
\end{eqnarray}

The non-abelian part of the Standard Model gauge algebra is engineered via toric methods (so-called ``tops'' \cite{Candelas:1996su, Bouchard:2003bu}).
In this case, we obtain five inequivalent tops that realize $\mathfrak{su}(3) \times \mathfrak{su}(2)$ in codimension one of the elliptic fibration \eqref{eq:hypersurface_two_u1s} \cite{Lin:2014qga}.
Furthermore, in each such top, we have the freedom of identifying the hypercharge $\fku(1)$ with a linear combination of the two geometrically realized $\fku(1)$s; the orthogonal combination then serves as the selection rule.
All such identifications compatible with the geometric spectrum have been listed in \cite{Lin:2014qga}, together with the possible dimension four and five operators of the Standard Model, which are and are not forbidden by the selection rule.

Again, one can attempt to find flux configurations that realize the chiral spectrum of the Standard Model.
For this fibration however, there is additional complexity arising from the fact that there are now additional matter curves which have the same representation under the Standard Model group, but differ by the charge under the selection rule $\fku(1)$.
Thus, there can be some ambiguity as to how to identify the geometrically realized states with those of the Standard Model.
Due to these ambiguities, it is tricky to find flux solutions that do not produce any chiral exotics.
With the techniques presented in \cite{Lin:2016vus}, the realization closest to the Standard Model spectrum is for a fibration over $B = \text{Bl}_1 \bbP^3$ and contains one chiral exotic pair of triplets and four singlets charged only under the selection rule $\fku(1)$.
In this realization, the $\fku(1)$ is of Peccei--Quinn type, i.e, the Higgs-up and -down doublet are charged differently.

While the selection rule does forbid certain dimension four operators, there are still some problematic ones left.
For example, the charge assignments are such that the Higgs-down and the lepton doublets have the same charges under the selection rule.
Therefore, any Higgs Yukawa coupling of down-type quarks also lead to lepton- and baryon-number violating operators involving two quarks and a lepton.
Furthermore, the selection rule $\fku(1)$ remains massless even in the presence of flux, and would need a different mechanism to lift the photons from the massless spectrum or to decouple them from the visible sector.

To circumvent these issues, one can instead use a discrete symmetry as selection rule.
As we will show now, such F-theory models can be constructed together with flux solutions that produce no chiral exotics, and with no problematic dimension four operators.

\subsection{An F-theory realization of matter parity}

As a final example of F-theory model building, we present a construction of the Standard Model with matter parity extension \cite{Cvetic:2018ryq}.
The technology for that only became available with the understanding of multi-section geometries.

In the previous section, we have discussed how a single abelian discrete gauge factor can be described in F-theory by a genus-one fibration.
However, for the Standard Model, we also need a $\fku(1)$, which naively requires the existence of rational sections.
One possible way to reconcile the two is to consider elliptic fibrations that have non-trivial Mordell--Weil groups and torsional cohomology.
However, the presence of terminal singularities there would then make the description of $G_4$-fluxes, at least in our current understanding, impossible.
Fortunately, it was realized in \cite{Klevers:2014bqa, Grimm:2015wda} that one can also use genus-one fibrations that have multiple \textit{independent} $n$-section classes.
In that case, they give rise in the dual M-theory compactification to multiple massless $\fku(1)$s, only one of which has to be identified with the linear combination of KK- and the massive $\fku(1)$.
The remaining $\fku(1)$s then can be uplifted to genuinely massless $\fku(1)$s in F-theory.

With realistic particle physics in mind, the simplest such fibration is again a toric hypersurface with fiber in a $\bbP^1 \times \bbP^1$ ambient space whose coordinates are $[x :t] \times [y:s]$.
In the defining polynomial,
\begin{align}
	(b_1\,y^2 + b_2\,s\,y + b_3\,s^2) \, x^2 + (b_5\,y^2 + b_6\,s\,y + b_7\,s^2)\,x\,t + (b_8\,y^2 + b_9\,s\,y + b_{10}\,s^2)\,t^2  \, ,
\end{align}
the two independent bisection classes are defined by the intersections with $\{x=0\}$ and $\{y=0\}$.
By choice, one identifies the KK/massive $\fku(1)$ with the divisor class $[x]$.
Then, the linear combination $[y] - [x] + \pi( ([y]-[x]) \cdot [x]) $, where the last term---the projection of the 4-cycle $([y] - [x]) \cdot [x]$ to the base---ensures the proper uplift to F-theory (compare to the Shioda map \eqref{eq:Shioda-map} in the case of rational sections).
Note that because the gauge symmetry is now $\fku(1) \times \bbZ_2$, there is no ambiguity in the identification of the hypercharge.
However, by identifying the $\bbZ_2$ symmetry as matter parity, there are two conventions of charge assignments which are physically equivalent (see \cite{Cvetic:2018ryq} and references therein).
Essentially, they differ by whether the left-handed quarks are charged odd or even under the $\bbZ_2$.

When we introduce the non-abelian gauge part with toric methods, we obtain the following geometrically realized spectrum:
\begin{align}
	\begin{split}
		& ({\bf 3,2})_{(\frac16,-)} \, , \quad \overline{\three}_{(-\frac23, +)} \, , \quad \overline{\three}_{(-\frac23, -)} \, , \quad \overline{\three}_{(\frac13, +)} \, , \quad \overline{\three}_{(\frac13, -)} \, , \\
		& \two_{(-\frac12, +)} \, , \quad \two_{(-\frac12, -)} \, , \quad \one_{(1,+)} \, , \quad \one_{(1,-)} \, , \quad \one_{(0,-)} \, .
	\end{split}
\end{align}
Because there is only one bifundamental state, its $\bbZ_2$ charge fixes the convention for the matter parity charges: all Standard Model fermions, i.e., the left-handed leptons and right-handed quarks and electrons must have odd $\bbZ_2$ charge.
The most phenomenologically appealing $G_4$ configuration therefore should induce chirality $\chi = 3$ for these states, whereas those states with even parity should have vanishing $\chi$.
Indeed, as demonstrated in \cite{Cvetic:2018ryq}, one can find, already on the simplest base $B = \bbP^3$, multiple such configurations.
These examples are the first F-theory constructions that reproduce the Standard Model spectrum at the chiral level, and has no problematic dimension four operators due to the presence of the matter parity selection rule.
As a final remark, note that this model also includes a singlet uncharged under the Standard Model gauge group, but is odd under parity.
Because it is a real representation, there cannot be any chirality associated with it (which is also ensured geometrically, see \cite{Mayrhofer:2014haa, Lin:2015qsa, Cvetic:2018ryq}).
Phenomenologically, it can be identified with right-handed neutrinos.

While the above models have the correct chiral spectrum, we cannot make a statement about the spectrum of vector-like pairs.
Since the Higgs doublets in the MSSM are vector-like, it would be interesting to apply the methods of \cite{Bies:2014sra, Bies:2017abs} to these models to obtain more realistic F-theory models of particle physics.

\section{Other Aspects of Abelian Symmetries in F-theory}\label{sec:outlook}

In these notes, we have primarily focused on the particle physics applications of abelian gauge symmetries in F-theory.
But of course, this does not do justice to the significant efforts that address other formal questions and applications.
In this last section, we will summarize and highlight some of the recent developments orthogonal to the model building aspect of abelian symmetries.

\subsubsection*{Anomalies and the Swampland}

One active subject can be motivated by the question about the upper bound of $\fku(1)$ charges in F-theory.
At the moment, explicit constructions have realized $q_\text{max}=4$ \cite{Raghuram:2017qut}, and it has been recently conjecture \cite{valandro_talk_F-theory18}---based on matrix factorization techniques and duality to type II \cite{Collinucci:2018aho}---that the upper bound is 6.
As shown in \cite{Taylor:2018khc}, there is no pure field theoretic arguments that would forbid higher charge states.
Hence, this conjecture can be interpreted as a swampland criterion, similar to the charge constraint \eqref{eq:charge_constraint_compact} related to the global gauge group structure.

The field theory arguments are based anomaly considerations, which are very stringent in 6D supergravity theories.
When we compactify F-theory to 4D, the anomaly conditions also depend on the $G_4$-flux, which have a geometric description, but are not ``geometrized'' by the elliptic fibration, i.e., the configuration needs to be specified in addition to the fibration (see appendix \ref{app:fluxes}).
However, one can reverse the logic and use anomaly cancellation to constrain the geometry of fourfolds.
Indeed, following the initial work \cite{Cvetic:2012xn}, it has been subsequently realized that a geometric reformulation of 4D gauge anomaly cancellation leads to certain geometric properties, which appear to be satisfied for all explicit model constructed so far in the literature \cite{Grimm:2015zea, Bies:2017fam,  Bies:2017abs, Corvilain:2017luj}.
Moreover, it has also been observed that discrete anomalies---in particular chiral anomalies associated with $\bbZ_n$ symmetries---of the 4D effective field theory are intimately related to the quantization condition of $G_4$ \cite{Lin:2015qsa, Lin:2016vus, Cvetic:2018ryq}.
So far though, there is no proof of these observations.

\subsubsection*{Heterotic duality and mirror symmetry}

While we have extensively used the duality to M-theory to explain the physics of F-theory compactifications, we have not touched upon the duality to the heterotic string \cite{Vafa:1996xn, Morrison:1996na, Morrison:1996pp}.
Under this duality, the fate of abelian symmetries on the heterotic side has been recently studied in \cite{Cvetic:2015uwu} and \cite{Cvetic:2016ner} (for continuous and discrete symmetries, respectively).
At a technical level, the analysis relied on a toric description of the so-called stable degeneration limit, which identifies the dual heterotic geometry and the gauge bundle data.

In the toric set-up, one stumbles across a surprising connection between abelian symmetries and mirror symmetry.
Concretely, consider a genus-one fibration $Y$ whose torus fibers $\mathfrak{f}$ are embedded into a toric ambient space ${\cal A}$.
One can then consider a fibration $Y'$ whose fibration is fibers $\mathfrak{f}'$ are mirror dual to $\mathfrak{f}$, and hence embedded into a toric ambient space ${\cal A}'$ that is the dual to ${\cal A}$.
It was first observed in \cite{Klevers:2014bqa} that if $Y$ has torsional Mordell--Weil group $\bbZ_n$, then the ``fiber-mirror-dual'' model $Y'$ is a  genus-one fibration with an $n$-section.
For F-theory purposes, one might therefore say that ``fiber-mirror-symmetry'' exchanges Mordell--Weil torsion with Tate--Shafarevich group.
This observation has been since further strengthened \cite{Oehlmann:2016wsb, Cvetic:2016ner}.
However, there are a few mirror dual pairs which do not seem to fit into this pattern.
To understand these examples, as well as a clearer physical picture of the phenomenon, additional efforts would be required.

\subsubsection*{Abelian symmetries in 6D SCFTs}

One of the recent achievements of F-theory is the classification of 6D ${\cal N} = (1,0)$ superconformal field theories (SCFTs) \cite{Heckman:2015bfa, Bhardwaj:2015xxa} using the geometry of elliptic fibrations (see \cite{Heckman:2018jxk} for a recent review).
Within this classification, only non-abelian gauge symmetries appear.
While this is consistent with field theory considerations, it was not until recently \cite{Lee:2018ihr} that it was understood how gauged $\fku(1)$s in compact F-theory geometries become global symmetries upon decoupling gravity.
Geometrically, the decoupling limit is where one takes the base $B$ to infinite volume.
In \cite{Lee:2018ihr}, it was shown that in this limit, the gauge coupling associated with the $\fku(1)$ always approaches zero, thus explaining the global nature of the symmetry.

In this context, discrete symmetries are much less understood.
For one, the geometric incarnation of the gauge coupling for such a symmetry has not been explored yet.
However, there are some evidence that discrete symmetries are important to distinguish certain strongly coupled sectors \cite{Anderson:2018heq}.
In these examples, the geometry are genus-one fibrations over compact bases which have so-called ``multiple fibers'' over singular points of the base.
Resolving these singularities reveal that the strongly coupled sector have additional singlets compared to models without multiple fibers (but singular points in base) \cite{DelZotto:2014fia}.
In the genus-one fibration, one can readily see that these singlets are charged only under the discrete symmetry related to the multi-section.
It would be interesting to analyze the decompactification limit of these models and explore if genus-one fibrations could add something new to the classification of 6D SCFTs.

\section*{Acknowledgements}

M.C.~would like to thank Igor Klebanov,  with whom she co-organized TASI 2017  School on ``Physics at the Fundamental Frontier'', for an enjoyable and constructive collaboration.
Furthermore, she thanks Senarath de Alwis, Oliver DeWolfe and especially Thomas DeGrand, for help and support during all stages of the TASI 2017 School organization. 
She is also grateful to all the lecturers for delivering excellent lectures and for inspiring atmosphere, and to all the participants for their dedication and involvement. 
L.L.~would like to thank the organizers of the workshop ``Geometry and Physics of F-theory'' (18w5190) at the Banff International Research Station.
We are both indebted to our collaborators on F-theory and related research: R.~Donagi, A.~Grassi, T.~Grimm, J.~Halverson, J.~Heckman, D.~Klevers, C.~Lawrie, M.~Liu, C.~Mayrhofer, P.~Oehleman, H.~Piragua, M.~Poretschkin,  P.~Song, W.~Taylor, O.~Till, T.~Weigand.
This work is supported by the DOE Award DE-SC0013528.
M.C.~further acknowledges the support by the Fay R.~and Eugene L.~Langberg Endowed Chair and the Slovenian Research Agency.

\appendix

\section{Gauge Fluxes and Chiral Spectra in F-theory}\label{app:fluxes}

While gauge fluxes are not directly related to abelian symmetries in F-theory, both of them are of global nature.
It is therefore not surprising that most of the work concerning global descriptions of gauge fluxes arose as an effort parallel to the understanding of $\fku(1)$s \cite{Krause:2011xj, Grimm:2011fx, Krause:2012yh, Collinucci:2010gz, Braun:2011zm, Collinucci:2012as, Cvetic:2012xn, Cvetic:2013uta, Borchmann:2013hta, Bies:2014sra, Braun:2014xka, Mayrhofer:2014haa, Lin:2015qsa, Lin:2016vus, Bies:2017fam, Bies:2017abs, Corvilain:2017luj, Cvetic:2018ryq}.
Because fluxes are an essential part of the examples presented in section \ref{sec:application}, it seems appropriate to include a brief introduction to the topic of fluxes, although we will have to refer to the review \cite{Weigand:2018rez} for more details and also appropriate references.

\subsection{Geometric description of gauge fluxes via duality to M-theory}

Our understanding of gauge fluxes arise from the M-/F-theory duality.
In M-theory compactified on a fourfold $Y$, on can turn on a background profile of the 3-form potential $C_3$ on the internal space.
Its field strength $G_4 = \text{d} C_3$ is then a closed 4-form, i.e., can be described by a cohomology form in $H^4(Y)$.
To preserve spacetime supersymmetry, the 4-form has to lie in $H^{2,2}(Y) \subset H^4(Y)$.
Under the assumption of the Hodge conjecture, such forms are always Poincar\'{e}-dual to algebraic 4-cycles.

A subset of algebraic 4-cycles are linear combinations of intersection products of divisors.
These span a subspace of $H^{2,2}$, called the primary vertical $(2,2)$-forms, or just vertical fluxes.
While there are other types of fluxes (the ``horizontal'' and the ``remainder'' pieces of $H^{2,2}$), the vertical ones are usually the only part relevant for the computation of the chiral spectrum in F-theory.
Now we have seen in section \ref{sec:u1} that the set of divisors of an elliptic fibration is completely captured by the Shioda--Tate--Wazir theorem \eqref{eq:ShiodaTateWazir}.
Likewise, it is conjectured that the same holds on genus-one fibrations by replacing the sections with independent multi-sections \cite{Braun:2014oya}.
Hence, given an explicit global model for which we know the full gauge symmetry, we can also systematically determine all vertical fluxes.
It is worth noting that the geometric description of fluxes in terms of 4-cycles is only possible on a smooth fourfold.
This means in particular that for F-theory models with discrete symmetries, a flux and chirality analysis with known methods is only possible on the associated multi-section geometry, whereas for the Jacobian fibration with its terminal singularities, new set of computational tools would be required.

So far, we have described fluxes in the M-theory set-up.
In order for them to uplift to F-theory, they have to satisfy some additional constraints.
The first set are the so-called transversality conditions, which in terms of the 4-cycle class $[G_4]$ of the flux can be phrased via intersection numbers:
\begin{align}\label{eq:transversality_condition_fluxes}
	\begin{split}
		[G_4] \cdot D^{(1)}_B \cdot D^{(2)}_B = [G_4] \cdot D^{(3)}_B \cdot Z \, .
	\end{split}
\end{align}
Here, $D_B^{(i)}$ are any divisors pulled-back from the base $B$.
Meanwhile, $Z$ denotes the divisor class of the embedding of the base into the full fibration; for an elliptic fibration this is simply the class of the zero section.
For a genus-one fibration, this is the class of the multi-section which is chosen as the divisor giving rise to the Kaluza--Klein $\fku(1)$, see section \ref{sec:discrete}.
Furthermore, in the presence of non-abelian gauge symmetries, a flux will generically break it unless it satisfies
\begin{align}
	[G_4] \cdot E_i \cdot D_B = 0 \, ,
\end{align}
for any pull-back divisor $D_B$ and any exceptional divisor $E_i$.

Finally, the flux has to satisfy the so-called quantization condition
\begin{align}
	G_4 + \frac{1}{2} c_2(Y) \in H^{2,2}(Y, \bbZ) = H^{2,2}(Y) \cap H^4(Y, \bbZ) \, ,
\end{align}
where $c_2(Y)$ is the second Chern class of the tangent bundle of $Y$.
This condition is notoriously difficult to check explicitly.
However, it has interesting consequences regarding certain topological quantities.
For example, a properly quantized flux must lead to an integer M2-/D3-tadpole
\begin{align}
	n_\text{D3} = \frac{\chi_e (Y)}{24} - \frac{1}{2} \int_Y G_4 \wedge G_4 \, ,
\end{align}
with $\chi_e$ the Euler number.
Furthermore, it has been observed recently that discrete anomalies such as Witten's $SU(2)$ anomaly or chiral anomalies of discrete symmetries are canceled if and only if the flux are properly quantized.

\subsection{Matter surfaces and chiral spectra}

To compute the chiral spectrum, we also need a geometric object associated with each matter representation in F-theory.
These are the so-called matter surfaces $\gamma_{\bf R}$, which are obtained by fibering codimension two fiber components $\Gamma_{\bf R}$ carrying weights of a representation ${\bf R}$ over the corresponding curve $C_{\bf R}$ on the base (recall that the base $B$ in this case is a threefold):
\begin{equation}
	\begin{tikzcd}[row sep = normal, column sep= normal  ]
		\Gamma \arrow[hook, r] & \gamma_{\bf R} \arrow[d] \\
		& C_{\bf R}
	\end{tikzcd} \, .
\end{equation}
As the name suggest, $\gamma_{\bf R}$ is complex surface, which is an algebraic 4-cycle that in almost all explicit examples turn out to be vertical.
Given a $G_4$-flux and its dual 4-cycle class $[G_4]$, the chiral index of matter in representation ${\bf R}$ is computed as
\begin{align}\label{eq:chirality}
	\chi( {\bf R}) = \int_{ \gamma_{\bf R}} G_4 = [G_4] \cdot [\gamma_{\bf R} ] \, ,
\end{align}
where $\cdot$ denotes the intersection product on the fourfold.

With suitable computational methods, the intersection number \eqref{eq:chirality} can be reduced to intersection numbers of divisors in the base.
For the examples presented in section \ref{sec:application}, these led to a general formula for the chiral indices of all matter representations which capture the full dependence on flux parameters and the fibration data over any base.
By varying these data and the choice of base $B$ of the fibration, one can then systematically scan for flux configurations that lead to desirable spectra.

Going beyond the chiral spectrum, it is also possible to determine the spectrum of vector-like pairs.
To determine these, however, requires more sophisticated methods and mathematical background, which have only been developed recently \cite{Bies:2014sra, Bies:2017fam}.
We again refer to \cite{Weigand:2018rez} for more details.

\bibliography{FTheory}{}
\bibliographystyle{JHEP} 

\end{document}